\documentclass[lettersize,journal]{IEEEtran}

\usepackage{amsmath,amsfonts,amssymb}
\usepackage{graphicx}
\usepackage[caption=false,font=normalsize,labelfont=sf,textfont=sf]{subfig}
\usepackage{textcomp}
\usepackage{stfloats}
\usepackage{url}
\usepackage{verbatim}
\usepackage{cite}
\usepackage{booktabs}
\usepackage{multirow}
\usepackage{rotating}
\usepackage{array}
\usepackage{makecell}
\usepackage{float}
\usepackage{amsthm}
\usepackage{booktabs}
\usepackage{multirow}
\usepackage{makecell}
\usepackage{graphicx}

\usepackage{algorithm}
\usepackage{algorithmic}

\usepackage{threeparttable}

\begin{document}

\title{A Unified Fractional Spectral Framework for Spatiotemporal Graph Signals: Bi-Fractional Transform and Geodesic Coupling}

\author{Mingzhi~Wang, Manjun~Cui, Feiyue~Zhao, Yangfan~He, and Zhichao~Zhang,~\IEEEmembership{Member,~IEEE}
\thanks{This work was supported in part by the Open Foundation of Hubei Key Laboratory of Applied Mathematics (Hubei University) under Grant HBAM202404; in part by the Foundation of Key Laboratory of System Control and Information Processing, Ministry of Education under Grant Scip20240121; and in part by the Startup Foundation for Introducing Talent of Nanjing Institute of Technology under Grant YKJ202214. \emph{(Corresponding author: Zhichao~Zhang.)}}
\thanks{Mingzhi~Wang, Manjun~Cui, and Feiyue~Zhao are with the School of Mathematics and Statistics, Nanjing University of Information Science and Technology, Nanjing 210044, China (e-mail: wmz200208@163.com; cmj1109@163.com; 202511150010@nuist.edu.cn).}
\thanks{Yangfan~He is with the School of Communication and Artificial Intelligence, School of Integrated Circuits, Nanjing Institute of Technology, Nanjing 211167, China, and also with the Jiangsu Province Engineering Research Center of IntelliSense Technology and System, Nanjing 211167, China (e-mail: Yangfan.He@njit.edu.cn).}
\thanks{Zhichao~Zhang is with the School of Mathematics and Statistics, Nanjing University of Information Science and Technology, Nanjing 210044, China, with the Hubei Key Laboratory of Applied Mathematics, Hubei University, Wuhan 430062, China, and also with the Key Laboratory of System Control and Information Processing, Ministry of Education, Shanghai Jiao Tong University, Shanghai 200240, China (e-mail: zzc910731@163.com).}}

\maketitle

\begin{abstract}
	Graph signal processing extends spectral analysis to data supported on irregular domains. Existing fractional transforms for two-dimensional graph signals, including the two-dimensional graph fractional Fourier transform (GFRFT), typically impose a shared fractional order across dimensions, which limits adaptivity to heterogeneous spatiotemporal spectra. To address this limitation, we propose the two-dimensional graph bi-fractional Fourier transform, which assigns independent fractional orders to the factor graphs of a Cartesian product, enabling decoupled spectral control while preserving separability, unitarity, and invertibility. To further resolve the basis ambiguity in temporal fractional analysis, we develop a geodesic-coupled GFRFT by constructing a coupling path along the principal geodesic on the unitary manifold, thereby unifying graph-induced and discrete temporal bases with guaranteed unitarity and a closed-form inverse. Building on these transforms, we derive a differentiable Wiener-type filtering framework with a hybrid optimization strategy: the fractional orders are learned end-to-end from data, while the coupling parameter is fixed as a structural regularizer. Experiments on real-world time-varying graph datasets and dynamic image restoration tasks demonstrate consistent gains over state-of-the-art fractional transforms and competitive learning-based baselines.
\end{abstract}


\bstctlcite{BSTcontrol}

\section{Introduction}

\begin{figure*}[!t]
	\centering
	\includegraphics[width=\textwidth]{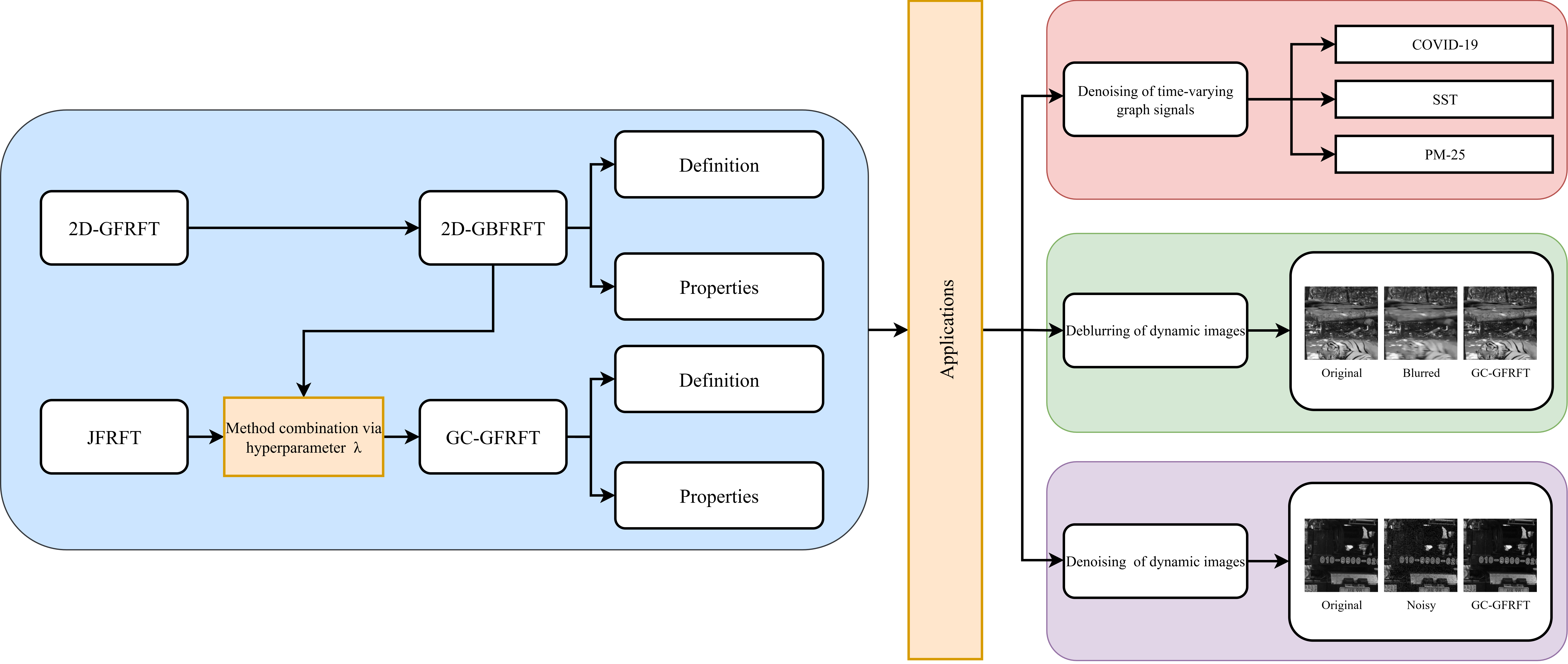}
	\caption{Overview of the proposed GC-GFRFT framework and its applications.}
	\label{fig:flow3}
\end{figure*}

Signals originating from real-world systems are often supported on complex networks or graphs, thereby exhibiting non-Euclidean topologies. Representative examples include spatiotemporal measurements from environmental sensing networks~\cite{1,2,3}, interactions in social networks~\cite{4,5}, brain connectivity graphs in neuroscience~\cite{6,7}, and pixel- or region-level relations in image analysis~\cite{8,9,fu2024fractional,yang2026refining}. The irregularity of these domains makes it difficult to directly apply classical techniques developed for Euclidean signals. Graph signal processing (GSP) provides a principled extension of traditional methodologies, including Fourier analysis~\cite{10,11,12,13,14}, wavelet transforms~\cite{15,16,17,18}, sampling theory~\cite{19,20,21}, and filtering~\cite{10,12,14,22,23,24}, to signals defined on graphs, enabling efficient modeling and analysis on irregular topologies.

In GSP, the graph Fourier transform (GFT) represents signals in a spectral domain induced by the eigendecomposition of a graph operator. Common choices of spectral operators include the graph Laplacian and the adjacency matrix. The Laplacian emphasizes smoothness and is well suited to diffusion-type processes, whereas the adjacency matrix emphasizes connectivity and can better capture nonstationary behavior and localized high-frequency patterns~\cite{10,11,14,25}. In this work, we adopt the adjacency matrix as the spectral operator to model signals with strong local variations. We focus on weighted undirected graphs so that the adjacency matrix admits an orthonormal eigendecomposition, yielding a unitary spectral basis. Despite its effectiveness, the GFT relies on fixed integer-order spectral modes, which limits its ability to adapt spectral representations to multiscale and heterogeneous structures.

To provide a continuous degree of freedom, the graph fractional Fourier transform (GFRFT) extends the GFT by introducing a fractional order that parameterizes a family of graph spectral transforms. Such representations have been applied to denoising, spectrum sparsification, and graph filtering~\cite{26,27,28,29,30,31,32}. In the classical fractional Fourier transform (FRFT), the fractional order is often interpreted via time-frequency rotation. This geometric interpretation does not directly carry over to graph settings because graph frequencies are induced by a chosen graph operator rather than Euclidean geometry. Accordingly, we treat the fractional order as a continuous spectral parameterization in the graph setting. Most existing GFRFT methods, however, focus on single-graph signals and provide limited support for multidimensional or multi-structure graph data.

Many applications naturally involve multidimensional graph signals, such as time-vertex or space-attribute data. Cartesian product graphs provide a convenient model for such joint structures by combining two factor graphs to represent multiway dependencies. Within this setting, the two-dimensional GFT (2D-GFT) performs spectral decompositions on the factor graphs and combines them through the Kronecker product, yielding a separable joint spectral representation with favorable computational properties~\cite{33,34,35}. Nonetheless, 2D-GFT remains restricted to integer-order descriptions. The two-dimensional GFRFT (2D-GFRFT) alleviates this limitation by applying a shared fractional order to both factor graphs~\cite{36}. This shared-order construction implicitly assumes that a single order can adequately control the two factor spectra. In strongly heterogeneous settings, such as when one dimension is relatively smooth while the other exhibits pronounced local variations, this constraint can significantly reduce expressiveness.

To address this limitation, we propose the two-dimensional graph bi-fractional Fourier transform (2D-GBFRFT). By assigning two independent fractional orders to the two factor graphs of a Cartesian product, the proposed bi-fractional design enables factor-wise tuning of the joint spectral coordinates. This improves modeling fidelity for heterogeneous two-dimensional graph signals while retaining the separable product-graph structure that supports efficient spectral-domain processing.

Beyond spatial heterogeneity, analyzing time-varying graph signals also requires selecting an appropriate temporal fractional basis. Joint spectral analysis is central to such signals, as exemplified by the joint time-vertex Fourier transform (JFT)~\cite{37,38} and the joint time-vertex FRFT (JFRFT)~\cite{39,40,41}. In practical spatiotemporal settings, the temporal dimension can be modeled either by a graph-induced fractional transform associated with a temporal graph, such as a path graph, or by a classical discrete FRFT (DFRFT) basis. Although both choices are typically unitary and thus energy preserving, they can induce different temporal spectral coordinate systems, which in turn affect subsequent filtering and representation.

This observation motivates our second contribution: a structure-preserving coupling mechanism for heterogeneous temporal fractional bases. A naive Euclidean interpolation between two unitary operators, such as linear mixing, generally breaks unitarity and can lead to ill-conditioned transforms, which compromise energy preservation and stable reconstruction~\cite{42,43}. Instead of interpolating in the ambient Euclidean space, we regard unitary operators as points on a Riemannian manifold and connect the two temporal endpoints through the principal geodesic~\cite{44}. The resulting one-parameter family, indexed by $\lambda\in[0,1]$, remains unitary for all $\lambda$, and its inverse is given by the Hermitian transpose. This intrinsic construction motivates the term geodesic-coupled, abbreviated as GC, and we refer to the resulting framework as the GC-GFRFT.

Wiener filtering is a fundamental mean-squared error (MSE) denoising technique~\cite{14,17,45,46,47}. Its graph counterpart applies spectral attenuation in the GFT domain and acts as a linear minimum-MSE estimator under stochastic priors~\cite{48,49}. To better align signal and noise spectra, GFRFT provides a tunable spectral domain. Early GFRFT-based filters derived their coefficients from prior statistical knowledge while using fixed fractional orders~\cite{45}. For time-varying signals, JFRFT-based filters employed grid search to optimize two hybrid fractional parameters, with a graph-induced basis used in the spatial domain and a discrete FRFT basis used in the temporal domain~\cite{39,40}. However, such discrete parameter sweeps become computationally expensive as the dimensionality increases.

Recent differentiable GFRFT formulations overcome this by enabling gradient-based learning of fractional orders and filter parameters~\cite{29}. More broadly, learning-based filtering frameworks are actively advancing graph signal processing~\cite{50}, leading to the exploration of learnable JFRFT designs for joint time-vertex spectral control~\cite{41}. These advances demonstrate that fractional-domain designs substantially enhance the expressiveness of spectral denoising and shaping on graphs.
Motivated by these advances and the fact that the temporal dimension admits two valid yet generally incompatible unitary fractional bases, we develop a differentiable Wiener-type filtering framework under a unified fractional spectral model. In our approach, the coupling parameter serves as a fixed structural regularizer selected via coarse search, while the fractional orders and diagonal spectral filters are learned end-to-end via gradient descent (GD).

In summary, we develop a unified fractional spectral framework for Cartesian product graphs and spatiotemporal graph signals. The main contributions are as follows:
\begin{itemize}
	\item We propose 2D-GBFRFT, a separable transform on Cartesian product graphs with independent fractional orders on the two factor graphs, enabling dimension-wise adaptivity and clarifying its distinction from JFRFT by using graph-induced bases in both dimensions.
	\item We introduce GC-GFRFT, a unitary geodesic coupling that bridges the graph-induced temporal basis and the DFRFT basis with a closed-form inverse, unifying 2D-GBFRFT ($\lambda=0$) and JFRFT ($\lambda=1$) as endpoint cases.
	\item We develop a regularized learnable Wiener-type filtering scheme that learns the fractional orders and diagonal spectral filters by GD while treating $\lambda$ as a fixed structural hyperparameter, and validate it on real data.
\end{itemize}

The remainder of this paper is organized as follows. Section~\ref{sec:prelim} reviews the preliminaries, including the fundamentals of GSP, Cartesian product graphs, and existing fractional transforms. Section~\ref{sec:2dgbfrft} proposes the 2D-GBFRFT, detailing its definition and theoretical properties. Section~\ref{sec:gcgfrft} introduces the GC-GFRFT, which resolves the temporal basis ambiguity via geodesic coupling on the unitary manifold. Section~\ref{sec:learnable_wiener} develops the differentiable Wiener filtering framework based on a hybrid optimization strategy. Section~\ref{sec:realdata} presents applications on real-world data, covering the denoising of time-varying graph signals and dynamic image restoration tasks. Finally, Section~\ref{sec:conclusion} concludes the paper and outlines future research directions, and technical proofs are provided in the Appendix. Fig.~\ref{fig:flow3} presents an overview of the proposed GC-GFRFT framework and its applications.

Notation: Bold lowercase letters denote vectors and bold uppercase letters denote matrices. The sets of real and complex numbers are denoted by $\mathbb{R}$ and $\mathbb{C}$, respectively. For a set $\mathcal{V}$, $|\mathcal{V}|$ denotes its cardinality. The operators $(\cdot)^*$, $(\cdot)^{\mathrm T}$, and $(\cdot)^{\mathrm H}$ represent the complex conjugate, transpose, and Hermitian transpose, respectively. The identity matrix of size $N$ is denoted by $\mathbf{I}_N$. The operator $\mathrm{diag}(\mathbf{h})$ denotes the diagonal matrix whose diagonal entries are given by $\mathbf{h}$. The operator $\mathrm{vec}(\cdot)$ denotes vectorization. The expectation operator is denoted by $\mathbb{E}\{\cdot\}$. The Euclidean ($\ell_2$) norm and Frobenius norm are denoted by $\|\cdot\|_2$ and $\|\cdot\|_F$, respectively. The Kronecker product, Kronecker sum, and Hadamard product are denoted by $\otimes$, $\oplus$, and $\circ$, respectively.

\section{PRELIMINARIES}
\label{sec:prelim}
\subsection{Graph Signals}
Let \( G = (\mathcal{V}, \mathbf{A}) \) be a weighted undirected graph, where \( \mathcal{V} = \{v_1, v_2, \dots, v_N\} \) denotes the node set with cardinality \( |\mathcal{V}| = N \), and \( \mathbf{A} \in \mathbb{R}^{N \times N} \) is the weighted adjacency matrix. An entry \( \mathbf{A}_{ij} > 0 \) indicates that an edge exists between nodes \( v_i \) and \( v_j \) with weight \( \mathbf{A}_{ij} \), while \( \mathbf{A}_{ij} = 0 \) indicates the absence of an edge. A graph signal is a function \( x: \mathcal{V} \to \mathbb{R} \) that assigns a scalar value \( x_i \) to each node \( v_i \in \mathcal{V} \). The values of the signal are collected into the vector \( \mathbf{x} = [x_1, x_2, \dots, x_N]^T \in \mathbb{R}^N \)~\cite{51}, which represents the signal in the vertex domain, with structure governed by the connectivity encoded in \( \mathbf{A} \).

To incorporate temporal dynamics, we assume that the graph signal is observed at \( T \) successive time instants with a unit sampling interval. Let \( \mathbf{x}_t \in \mathbb{R}^N \) denote the signal at time \( t \). The time-varying graph signal is represented by the matrix \( \mathbf{X} = [\mathbf{x}_1, \mathbf{x}_2, \dots, \mathbf{x}_T] \in \mathbb{R}^{N \times T} \). Both \( \mathbf{X} \) and its vectorized form \( \mathbf{x} = \mathrm{vec}(\mathbf{X}) \in \mathbb{R}^{NT} \) are referred to as the time-vertex signal~\cite{52,53}.

A Cartesian product \( G_1 \square G_2 \) of two weighted graphs \( G_1 = (\mathcal{V}_1, \mathcal{E}_1, \mathbf{A}_1) \) and \( G_2 = (\mathcal{V}_2, \mathcal{E}_2, \mathbf{A}_2) \) is a graph with vertex set \( \mathcal{V}_1 \times \mathcal{V}_2 \). Its edge set \( \mathcal{E} \) is defined by
\begin{align*}
	\{(i_1, i_2), (j_1, j_2)\} \in \mathcal{E} \iff & \left[\{i_1, j_1\} \in \mathcal{E}_1,\, i_2 = j_2\right] \\
	& \text{or}\ [i_1 = j_1,\, \{i_2, j_2\} \in \mathcal{E}_2],
\end{align*}
and the edge weight, corresponding to the entries of the adjacency matrix \( \mathbf{A} \) of \( G_1 \square G_2 \), is given by
\[
[\mathbf{A}]_{ (i_1, i_2), (j_1, j_2) } 
= [\mathbf{A}_1]_{i_1, j_1}\,\delta(i_2, j_2) + \delta(i_1, j_1)\,[\mathbf{A}_2]_{i_2, j_2}.
\]
Graphs \( G_1 \) and \( G_2 \) are referred to as the factor graphs of \( G_1 \square G_2 \).

Let \( G_n \) with vertex set \( \mathcal{V}_n = \{1,2,\dots,N_n\} \) have adjacency, degree, and Laplacian matrices \( \mathbf{A}_n \), \( \mathbf{D}_n \), and \( \mathbf{L}_n \), respectively, for \( n = 1, 2 \). When the vertices of the Cartesian product graph \( \mathcal{V}_1 \times \mathcal{V}_2 \) are ordered lexicographically, i.e., \( (1,1), (1,2), \dots, (N_1,N_2) \), the adjacency, degree, and Laplacian matrices of \( G_1 \square G_2 \) can be written as \( \mathbf{A}_1 \oplus \mathbf{A}_2 \), \( \mathbf{D}_1 \oplus \mathbf{D}_2 \), and \( \mathbf{L}_1 \oplus \mathbf{L}_2 \), respectively~\cite{33}.

\subsection{2D-GFRFT}

We consider the adjacency matrix \( \mathbf{A}_n \) of a weighted graph \( G_n \). 
For real symmetric weighted graphs, \( \mathbf{A}_n \) is diagonalizable with eigenvalue decomposition 
\( \mathbf{\Lambda}_{A_n} = \mathbf{V}_{A_n}^{\mathrm{H}} \mathbf{A}_n \mathbf{V}_{A_n} \), 
where \( \mathbf{V}_{A_n} \) contains the eigenvectors of \( \mathbf{A}_n \) and \( \mathbf{\Lambda}_{A_n} \) is a diagonal matrix of eigenvalues.
The GFT matrix is defined as \( \mathbf{F}_{G_n} = \mathbf{V}_{A_n}^{\mathrm{H}} \), 
and the inverse is given by \( \mathbf{F}_{G_n}^{-1} = \mathbf{V}_{A_n} \)~\cite{11}. 

We define the GFRFT as
\begin{equation}
	\mathbf{F}_{G_n}^{\alpha_n} = \mathbf{V}_{F_n} \, \mathbf{J}_{F_n}^{\alpha_n} \, \mathbf{V}_{F_n}^{-1},
	\label{eq:gfrft_def}
\end{equation}
where \( \mathbf{J}_{F_n}^{\alpha_n} \) represents the fractional matrix in the graph Fourier domain, and  $\mathbf{V}_{F_n}$ is the eigenvector matrix of $\mathbf{F}_{G_n}$.
If \( \mathbf{F}_{G_n} \) is unitary, then \( (\mathbf{F}_{G_n}^{\alpha_n})^{\mathrm{H}} = \mathbf{F}_{G_n}^{-\alpha_n} \).

Given a graph signal $X: V_1 \times V_2 \to \mathbb{R}$ on the Cartesian product graph $G_1 \square G_2$,
let $N_i \triangleq |V_i|$ ($i=1,2$) and let $\mathbf{X}\in\mathbb{R}^{N_1\times N_2}$ denote the vertex-domain matrix representation of $X$.

\textit{Definition 1:}
The 2D-GFRFT of $\mathbf{X}$ is defined as
\begin{equation}
	\mathbf{X}_f^{\alpha}
	= \mathbf{F}_{G_1}^{\alpha}\,\mathbf{X}\,\big(\mathbf{F}_{G_2}^{\alpha}\big)^{\mathrm T},
	\label{eq:2dgfrft_matrix}
\end{equation}
where $\mathbf{F}_{G_i}^{\alpha}$ ($i=1,2$) are the $\alpha$-order GFRFT matrices of $G_i$.
The inverse transform is
\begin{equation}
	\mathbf{X}
	= \mathbf{F}_{G_1}^{-\alpha}\,\mathbf{X}_f^{\alpha}\,\big(\mathbf{F}_{G_2}^{-\alpha}\big)^{\mathrm T}.
	\label{eq:2dgfrft_matrix_inv}
\end{equation}

Equivalently, in vectorized form, letting
\( \mathbf{x} \triangleq \mathrm{vec}(\mathbf{X}) \) and
\( \mathbf{x}_f^{\alpha} \triangleq \mathrm{vec}(\mathbf{X}_f^{\alpha}) \), we have
\begin{equation}
	\mathbf{x}_f^{\alpha} = \big(\mathbf{F}_{G_2}^{\alpha} \otimes \mathbf{F}_{G_1}^{\alpha}\big)\,\mathbf{x}
	\;\triangleq\; \mathbf{F}_{2D}^{\alpha}\,\mathbf{x},
	\label{eq:2dgfrft_vec}
\end{equation}
with the corresponding inverse
\begin{equation}
	\mathbf{x} = \big(\mathbf{F}_{G_2}^{-\alpha} \otimes \mathbf{F}_{G_1}^{-\alpha}\big)\,\mathbf{x}_f^{\alpha}
	\;\triangleq\; \mathbf{F}_{2D}^{-\alpha}\,\mathbf{x}_f^{\alpha}.
	\label{eq:2dgfrft_vec_inv}
\end{equation}

The 2D-GFRFT thus provides a joint spectral representation spanned by the fractional eigenmodes of the two factor graphs and reduces to the classical 2D-GFT when \( \alpha = 1 \)~\cite{36}.

\subsection{JFRFT}

The FRFT extends the classical Fourier transform by introducing a continuous order parameter \( \alpha \)~\cite{54,55,56}. Its discrete version, known as the DFRFT, is represented by a matrix \( \mathbf{F}^{\alpha} \in \mathbb{C}^{T \times T} \), which reduces to the discrete Fourier transform (DFT) when \( \alpha = 1 \)~\cite{57}.

JFRFT generalizes this concept to time-varying graph signals, incorporating both the graph structure and temporal dynamics. Let \( \mathbf{X} \in \mathbb{C}^{N \times T} \) denote a signal matrix defined over a graph with \( N \) vertices and \( T \) time instants, and let \( \mathbf{x} = \mathrm{vec}(\mathbf{X}) \in \mathbb{C}^{NT} \) be its vectorized form.

\textit{Definition 2:} 
The JFRFT of $\mathbf{X}$ is defined as
\begin{equation}
	\mathbf{X}_{J}^{(\alpha,\beta)}
	= \mathbf{F}_{\mathrm G}^{\beta}\, \mathbf{X}\, \big(\mathbf{F}^{\alpha}\big)^{\mathrm T},
	\label{eq:jfrft_matrix}
\end{equation}
where $\mathbf{F}^{\alpha}$ is the $\alpha$-order DFRFT matrix, 
$\mathbf{F}_{\mathrm G}^{\beta}$ is the $\beta$-order GFRFT matrix, 
and $\mathbf{X}_{J}^{(\alpha,\beta)}$ denotes the JFRFT-domain representation of $\mathbf{X}$. 
The inverse transform is
\begin{equation}
	\mathbf{X}
	= \mathbf{F}_{\mathrm G}^{-\beta}\, \mathbf{X}_{J}^{(\alpha,\beta)}\, \big(\mathbf{F}^{-\alpha}\big)^{\mathrm T}.
	\label{eq:jfrft_matrix_inv}
\end{equation}

Equivalently, in vectorized form, we have
\begin{equation}
	\mathbf{x}_{J}^{(\alpha,\beta)}
	= \big(\mathbf{F}^{\alpha} \otimes \mathbf{F}_{\mathrm G}^{\beta}\big)\,\mathbf{x}
	\;\triangleq\; \mathbf{F}_{J}^{(\alpha,\beta)}\,\mathbf{x},
	\label{eq:jfrft_vec}
\end{equation}
with the corresponding inverse
\begin{equation}
	\mathbf{x}
	= \big(\mathbf{F}^{-\alpha} \otimes \mathbf{F}_{\mathrm G}^{-\beta}\big)\,\mathbf{x}_{J}^{(\alpha,\beta)}
	\;\triangleq\; \mathbf{F}_{J}^{(-\alpha,-\beta)}\,\mathbf{x}_{J}^{(\alpha,\beta)}.
	\label{eq:jfrft_vec_inv}
\end{equation}

The JFRFT thus provides a joint fractional spectral representation of time-varying graph signals and reduces to the classical JFT when $\alpha=1$ and $\beta=1$.

\section{Two-Dimensional Graph Bi-Fractional Fourier Transform}
\label{sec:2dgbfrft}
\subsection{Definition}

The conventional 2D-GFRFT applies an identical fractional order to both factor graphs of a Cartesian product graph~\cite{36}. Although this uniform specification simplifies the transform, it may fail to capture structural heterogeneity between the two factors. In particular, when the factors exhibit significantly different characteristics, such as in space-time graph scenarios, a single shared order can introduce spectral aliasing or obscure directional information.

To overcome these limitations, we propose the 2D-GBFRFT. Unlike the conventional formulation, 2D-GBFRFT assigns two independent fractional orders, \( \alpha_1 \) and \( \alpha_2 \), to the factor graphs, enabling adaptivity along each dimension. This bi-fractional design improves both the flexibility and fidelity of spectral representations, particularly when the factor graphs have distinct structural properties.

Let two graphs be given as $G_1 = (\mathcal{V}_1, \mathcal{E}_1, \mathbf{A}_1)$ and $G_2 = (\mathcal{V}_2, \mathcal{E}_2, \mathbf{A}_2)$, with $N_1=|\mathcal{V}_1|$ and $N_2=|\mathcal{V}_2|$. Their adjacency matrices $\mathbf{A}_1$ and $\mathbf{A}_2$ admit the eigendecompositions
\begin{equation}
	\mathbf{A}_1 = \mathbf{V}_{A_1}\boldsymbol{\Lambda}_1 \mathbf{V}_{A_1}^{\mathrm{T}},
	\quad
	\mathbf{A}_2 = \mathbf{V}_{A_2}\boldsymbol{\Lambda}_2 \mathbf{V}_{A_2}^{\mathrm{T}}.
	\label{eq:2dgbfrft_lap}
\end{equation}
Given two independent fractional orders $\alpha_1, \alpha_2 \in \mathbb{R}$, the bi-fractional transform matrices are defined as
\begin{equation}
	\mathbf{F}_{G_1}^{\alpha_1} = \mathbf{V}_{A_1}\boldsymbol{\Lambda}_1^{\alpha_1}\mathbf{V}_{A_1}^{\mathrm{T}},
	\quad
	\mathbf{F}_{G_2}^{\alpha_2} = \mathbf{V}_{A_2}\boldsymbol{\Lambda}_2^{\alpha_2}\mathbf{V}_{A_2}^{\mathrm{T}}.
	\label{eq:2dgbfrft_fg}
\end{equation}

Given a graph signal $X: \mathcal{V}_1 \times \mathcal{V}_2 \to \mathbb{R}$ on the Cartesian product graph $G_1 \square G_2$, let $N_i \triangleq |\mathcal{V}_i|$ ($i=1,2$) and let $\mathbf{X}\in\mathbb{R}^{N_1\times N_2}$ denote the vertex-domain matrix representation of $X$.

\textit{Definition 3:}
The 2D-GBFRFT of $\mathbf{X}$ is defined as
\begin{equation}
	\mathbf{X}_f^{\alpha_1,\alpha_2}
	= \mathbf{F}_{G_1}^{\alpha_1}\,\mathbf{X}\,\big(\mathbf{F}_{G_2}^{\alpha_2}\big)^{\mathrm T},
	\label{eq:2dgbfrft_matrix}
\end{equation}
where $\mathbf{F}_{G_i}^{\alpha_i}$ ($i=1,2$) denote the $\alpha_i$-order GFRFT matrices of $G_i$, and $\mathbf{X}_f^{\alpha_1,\alpha_2}$ is the 2D-GBFRFT representation of $\mathbf{X}$.
The inverse transform is
\begin{equation}
	\mathbf{X}
	= \mathbf{F}_{G_1}^{-\alpha_1}\,\mathbf{X}_f^{\alpha_1,\alpha_2}\,\big(\mathbf{F}_{G_2}^{-\alpha_2}\big)^{\mathrm T}.
	\label{eq:2dgbfrft_matrix_inv}
\end{equation}

In vectorized form, letting $\mathbf{x}\triangleq\mathrm{vec}(\mathbf{X})$ and $\mathbf{x}_f^{\alpha_1,\alpha_2}\triangleq\mathrm{vec}(\mathbf{X}_f^{\alpha_1,\alpha_2})$, we have
\begin{equation}
	\mathbf{x}_f^{\alpha_1,\alpha_2}
	= \big(\mathbf{F}_{G_2}^{\alpha_2}\otimes \mathbf{F}_{G_1}^{\alpha_1}\big)\,\mathbf{x}
	\;\triangleq\; \mathbf{F}_{2D}^{(\alpha_1,\alpha_2)}\,\mathbf{x},
	\label{eq:2dgbfrft_vec}
\end{equation}
with the corresponding inverse
\begin{equation}
	\mathbf{x}
	= \big(\mathbf{F}_{G_2}^{-\alpha_2}\otimes \mathbf{F}_{G_1}^{-\alpha_1}\big)\,\mathbf{x}_f^{\alpha_1,\alpha_2}
	\;\triangleq\; \mathbf{F}_{2D}^{(-\alpha_1,-\alpha_2)}\,\mathbf{x}_f^{\alpha_1,\alpha_2}.
	\label{eq:2dgbfrft_vec_inv}
\end{equation}

\subsection{Properties}

The proposed 2D-GBFRFT inherits and extends several classical properties of FRFT and GFRFT. 
We summarize the main theoretical properties that establish its effectiveness in graph signal processing.

\textit{Property 1 (Identity):}  
If $\alpha_1 = \alpha_2 = 0$, the 2D-GBFRFT reduces to the identity transform, i.e.,
\begin{equation}
\mathbf{F}_{2D}^{(0,0)} 
= \mathbf{F}_{G_2}^{0} \otimes \mathbf{F}_{G_1}^{0} 
= \mathbf{I}_{N_2} \otimes \mathbf{I}_{N_1} 
= \mathbf{I}_{N_1N_2}.
\label{eq:2dgbfrft_identity}
\end{equation}

\textit{Property 2 (Reduction to 2D-GFRFT):}  
When $\alpha_1 = \alpha_2 = \alpha$, the 2D-GBFRFT reduces to the standard 2D-GFRFT:
\begin{equation}
\mathbf{F}_{2D}^{(\alpha,\alpha)} 
= \mathbf{F}_{G_2}^{\alpha} \otimes \mathbf{F}_{G_1}^{\alpha} 
= \mathbf{F}_{2D}^{\alpha}.
\label{eq:2dgbfrft_reduction}
\end{equation}

\textit{Property 3 (Unitarity):}  
If both $\mathbf{F}_{G_1}^{\alpha_1}$ and $\mathbf{F}_{G_2}^{\alpha_2}$ are unitary, 
then $\mathbf{F}_{2D}^{(\alpha_1,\alpha_2)}$ is also unitary. Specifically,
{\allowdisplaybreaks
	\begin{align}
		\big(\mathbf{F}_{2D}^{(\alpha_1,\alpha_2)}\big)^{\mathrm H}\,
		\mathbf{F}_{2D}^{(\alpha_1,\alpha_2)}
		&= \big(\mathbf{F}_{G_2}^{\alpha_2} \otimes \mathbf{F}_{G_1}^{\alpha_1}\big)^{\mathrm H}
		\big(\mathbf{F}_{G_2}^{\alpha_2} \otimes \mathbf{F}_{G_1}^{\alpha_1}\big) \notag\\
		&= \left( \bigl(\mathbf{F}_{G_2}^{\alpha_2}\bigr)^{\mathrm H}\mathbf{F}_{G_2}^{\alpha_2} \right)
		\ \otimes\
		\left( \bigl(\mathbf{F}_{G_1}^{\alpha_1}\bigr)^{\mathrm H}\mathbf{F}_{G_1}^{\alpha_1} \right) \notag\\
		&= \mathbf{I}_{N_1N_2},
		\label{eq:2dgbfrft_unitarity}
	\end{align}%
}
which establishes unitarity.

\textit{Property 4 (Index additivity):}  
For any two sets of orders $(\alpha_1,\alpha_2)$ and $(\beta_1,\beta_2)$, the additivity property holds:
\newpage
\begin{align}
	\mathbf{F}_{2D}^{(\alpha_1,\alpha_2)} \mathbf{F}_{2D}^{(\beta_1,\beta_2)}
	&= (\mathbf{F}_{G_2}^{\alpha_2} \otimes \mathbf{F}_{G_1}^{\alpha_1})
	(\mathbf{F}_{G_2}^{\beta_2} \otimes \mathbf{F}_{G_1}^{\beta_1}) \notag \\
	&= (\mathbf{F}_{G_2}^{\alpha_2}\mathbf{F}_{G_2}^{\beta_2})
	\otimes
	(\mathbf{F}_{G_1}^{\alpha_1}\mathbf{F}_{G_1}^{\beta_1}) \notag \\
	&= \mathbf{F}_{G_2}^{\alpha_2+\beta_2} \otimes \mathbf{F}_{G_1}^{\alpha_1+\beta_1} \notag \\
	&= \mathbf{F}_{2D}^{(\alpha_1+\beta_1,\,\alpha_2+\beta_2)}.
	\label{eq:2dgbfrft_additivity}
\end{align}
In particular, setting $(\beta_1,\beta_2) = (-\alpha_1,-\alpha_2)$ yields
\begin{equation}
	\mathbf{F}_{2D}^{(\alpha_1,\alpha_2)} 
	\mathbf{F}_{2D}^{(-\alpha_1,-\alpha_2)}
	= \mathbf{F}_{2D}^{(0,0)} 
	= \mathbf{I}_{N_1N_2},
	\label{eq:2dgbfrft_invertibility}
\end{equation}
which confirms the invertibility of the 2D-GBFRFT.

\section{Geodesic-Coupled Graph Fractional Fourier Transform}
\label{sec:gcgfrft}

\subsection{Definition}

Following the Cartesian product setting introduced earlier, we consider a spatiotemporal signal
$\mathbf{X}\in\mathbb{C}^{N_1\times N_2}$ supported on $G_1 \square G_2$, where $G_1$ is the spatial factor and $G_2$ is the temporal factor.
Let $\mathbf{F}_{G_1}^{\alpha}\in\mathbb{C}^{N_1\times N_1}$ denote the spatial GFRFT of order $\alpha\in\mathbb{R}$ on $G_1$.
Along the temporal dimension, we consider two unitary fractional transforms of the same order $\beta\in\mathbb{R}$:
a graph-induced temporal GFRFT $\mathbf{F}_{G_2}^{\beta}\in\mathbb{C}^{N_2\times N_2}$ and a discrete FRFT matrix
$\mathbf{F}^{\beta}\in\mathbb{C}^{N_2\times N_2}$, satisfying
\begin{equation}
	(\mathbf{F}_{G_2}^{\beta})^{\mathrm H}\mathbf{F}_{G_2}^{\beta}=\mathbf{I}_{N_2},
	\qquad
	(\mathbf{F}^{\beta})^{\mathrm H}\mathbf{F}^{\beta}=\mathbf{I}_{N_2}.
	\label{eq:gc_unitary_bases}
\end{equation}

Although both operators are unitary, they generally yield different temporal spectral coordinates. As discussed in the motivation,
a naive Euclidean interpolation between the two endpoints does not preserve unitarity in general.
We therefore connect them through a unitary geodesic on the unitary group $U(N_2)$, constructed via the principal matrix logarithm and exponential.

Define the relative change-of-basis operator as
\begin{equation}
	\mathbf{W}_t \triangleq \big(\mathbf{F}_{G_2}^{\beta}\big)^{\mathrm H}\mathbf{F}^{\beta}\in\mathbb{C}^{N_2\times N_2}.
	\label{eq:gc_Wt}
\end{equation}
Since both factors are unitary, $\mathbf{W}_t$ is unitary, and all its eigenvalues lie on the unit circle.

\textit{Assumption 1 (Principal logarithm well-defined).}
We assume $-1 \notin \sigma(\mathbf{W}_t)$ so that the principal matrix logarithm $\log(\mathbf{W}_t)$ is well defined and unique.

\textit{Definition 4 (Temporal geodesic coupling).}
For $\lambda\in[0,1]$, define
\begin{equation}
	\mathbf{F}_{t,\mathrm{GC}}^{(\lambda;\beta)}
	\triangleq
	\mathbf{F}_{G_2}^{\beta}\,
	\exp\!\big(\lambda\,\log(\mathbf{W}_t)\big).
	\label{eq:gc_Ft}
\end{equation}
This curve lies on $U(N_2)$ for all $\lambda$ and admits a closed-form inverse.

\textit{Equivalent phase-domain form:}
Because $\mathbf{W}_t$ is unitary, it admits an eigendecomposition
\begin{equation}
	\mathbf{W}_t=\mathbf{S}_t\mathbf{\Lambda}_t\mathbf{S}_t^{\mathrm H}, \;
	\mathbf{\Lambda}_t=\mathrm{diag}\big(e^{\mathrm j\theta_1},\ldots,e^{\mathrm j\theta_{N_2}}\big), \;
	\theta_k\in(-\pi,\pi),
	\label{eq:gc_Wt_eig}
\end{equation}
where $\mathbf{S}_t$ is unitary and $\theta_k$ are principal phases. Substituting \eqref{eq:gc_Wt_eig} into \eqref{eq:gc_Ft} yields
\begin{equation}
	\mathbf{F}_{t,\mathrm{GC}}^{(\lambda;\beta)}
	=
	\mathbf{F}_{G_2}^{\beta}\,
	\mathbf{S}_t\,\mathrm{diag}\big(e^{\mathrm j\lambda\theta_1},\ldots,e^{\mathrm j\lambda\theta_{N_2}}\big)\,
	\mathbf{S}_t^{\mathrm H},
	\label{eq:gc_phase}
\end{equation}
which makes explicit that $\lambda$ linearly interpolates the eigenphases of $\mathbf{W}_t$. For a fixed $\beta$, when $\mathbf{S}_t$ and $\{\theta_k\}_{k=1}^{N_2}$ are computed once, \eqref{eq:gc_phase} also enables efficient evaluation for multiple values of $\lambda$,
since varying $\lambda$ only updates the diagonal phase factors.

\textit{Definition 5 (GC-GFRFT).}
For $\mathbf{X}\in\mathbb{C}^{N_1\times N_2}$, define the GC-GFRFT as
\begin{equation}
	\mathbf{X}_{\mathrm{GC}}^{(\lambda;\alpha,\beta)}
	=
	\mathbf{F}_{G_1}^{\alpha}\,\mathbf{X}\,
	\big(\mathbf{F}_{t,\mathrm{GC}}^{(\lambda;\beta)}\big)^{\mathrm T}.
	\label{eq:gc_matrix}
\end{equation}
Equivalently, letting $\mathbf{x}\triangleq\mathrm{vec}(\mathbf{X})$ and
$\mathbf{x}_{\mathrm{GC}}^{(\lambda;\alpha,\beta)}\triangleq\mathrm{vec}\big(\mathbf{X}_{\mathrm{GC}}^{(\lambda;\alpha,\beta)}\big)$, the standard
$\mathrm{vec}$--Kronecker identity gives
\begin{equation}
	\mathbf{x}_{\mathrm{GC}}^{(\lambda;\alpha,\beta)}
	=
	\Big(\mathbf{F}_{t,\mathrm{GC}}^{(\lambda;\beta)}\otimes \mathbf{F}_{G_1}^{\alpha}\Big)\mathbf{x}
	\;\triangleq\;
	\mathbf{F}_{\mathrm{GC}}^{(\lambda;\alpha,\beta)}\mathbf{x}.
	\label{eq:gc_vec}
\end{equation}

\subsection{Properties}

We next study the global GC operator
\begin{equation}
	\mathbf{F}_{\mathrm{GC}}^{(\lambda;\alpha,\beta)}
	\triangleq
	\mathbf{F}_{t,\mathrm{GC}}^{(\lambda;\beta)}\otimes \mathbf{F}_{G_1}^{\alpha}
	\in\mathbb{C}^{(N_1N_2)\times(N_1N_2)}.
	\label{eq:FGC_def}
\end{equation}
Unless otherwise stated, all results hold under Assumption~1 and the unitarity of
$\mathbf{F}_{G_1}^{\alpha}$, $\mathbf{F}_{G_2}^{\beta}$, and $\mathbf{F}^{\beta}$.

\textit{Property 5 (Degeneracy to two endpoints).}
For $\lambda\in[0,1]$, the GC operator degenerates to
\begin{equation}
	\mathbf{F}_{\mathrm{GC}}^{(0;\alpha,\beta)}
	=
	\mathbf{F}_{G_2}^{\beta}\otimes \mathbf{F}_{G_1}^{\alpha},
	\qquad
	\mathbf{F}_{\mathrm{GC}}^{(1;\alpha,\beta)}
	=
	\mathbf{F}^{\beta}\otimes \mathbf{F}_{G_1}^{\alpha}.
	\label{eq:FGC_degeneracy}
\end{equation}

\textit{Proof:} See Appendix~\ref{app:proof_properties_5_7}.

\textit{Property 6 (Unitarity).}
For any $\lambda\in[0,1]$,
\begin{equation}
	\big(\mathbf{F}_{\mathrm{GC}}^{(\lambda;\alpha,\beta)}\big)^{\mathrm H}
	\mathbf{F}_{\mathrm{GC}}^{(\lambda;\alpha,\beta)}
	=
	\mathbf{I}_{N_1N_2}.
	\label{eq:FGC_unitary}
\end{equation}

\textit{Proof:} See Appendix~\ref{app:proof_properties_5_7}.

\textit{Property 7 (Endpoint symmetry).}
Define the swapped-endpoint temporal coupling
\begin{equation}
	\widetilde{\mathbf{F}}_{t,\mathrm{GC}}^{(\lambda;\beta)}
	\triangleq
	\mathbf{F}^{\beta}\exp\!\Big(\lambda\log\!\big((\mathbf{F}^{\beta})^{\mathrm H}\mathbf{F}_{G_2}^{\beta}\big)\Big),
	\label{eq:Ft_swap}
\end{equation}
and the corresponding global operator
\(
\widetilde{\mathbf{F}}_{\mathrm{GC}}^{(\lambda;\alpha,\beta)}
\triangleq
\widetilde{\mathbf{F}}_{t,\mathrm{GC}}^{(\lambda;\beta)}\otimes \mathbf{F}_{G_1}^{\alpha}.
\)
Then, for all $\lambda\in[0,1]$,
\begin{equation}
	\widetilde{\mathbf{F}}_{\mathrm{GC}}^{(\lambda;\alpha,\beta)}
	=
	\mathbf{F}_{\mathrm{GC}}^{(1-\lambda;\alpha,\beta)}.
	\label{eq:FGC_symmetry}
\end{equation}

\textit{Proof:} See Appendix~\ref{app:proof_properties_5_7}.
\section{Learnable Wiener Filtering in Fractional Spectral Domains}
\label{sec:learnable_wiener}

We extend the Wiener filtering principle to spatiotemporal graph signals under the proposed GC-GFRFT framework. Consider the stochastic observation model over a space--time domain
\begin{equation}
	\mathbf{Y} = \mathbf{G}_{S} \mathbf{X} \mathbf{G}_{T} + \mathbf{N},
	\label{eq:obs_model_matrix_gc}
\end{equation}
where $\mathbf{G}_{S} \in \mathbb{C}^{N_1 \times N_1}$ and $\mathbf{G}_{T} \in \mathbb{C}^{N_2 \times N_2}$ denote known linear system responses associated with the spatial graph $G_1$ and the temporal dimension, respectively. The matrix $\mathbf{X} \in \mathbb{C}^{N_1 \times N_2}$ represents the stochastic space--time signal, and $\mathbf{N}$ denotes additive noise. Letting $\mathbf{x}=\mathrm{vec}(\mathbf{X})$ and $\mathbf{n}=\mathrm{vec}(\mathbf{N})$, the model can be written in vectorized form as
\begin{equation}
	\mathbf{y} = \big(\mathbf{G}_{T}^{\mathsf{T}} \otimes \mathbf{G}_{S}\big)\mathbf{x} + \mathbf{n}.
	\label{eq:obs_model_vector_gc}
\end{equation}
When $\lambda=0$, the temporal GC basis reduces to the graph-induced temporal fractional basis, and the resulting transform specializes to the decoupled 2D-GBFRFT with independent spatial and temporal orders.

\subsection{Learnable Spectral Filtering via Gradient Descent}

A direct grid search over fractional orders incurs exponentially many configurations as the number of tunable parameters increases. To avoid this, we leverage the differentiability of the GC-GFRFT operator and jointly optimize the fractional orders and diagonal spectral filter coefficients via gradient-based learning.

Specifically, the fractional orders $(\alpha,\beta)$ are treated as learnable parameters, whereas the coupling parameter $\lambda$ is treated as a fixed structural hyperparameter that controls the geodesic coupling between the graph-induced temporal basis and the discrete temporal basis. In practice, $\lambda$ is selected by a coarse search with step size $0.1$ and then kept fixed during optimization; hence, $\lambda$ is excluded from gradient updates.

\textit{Theorem 1 (Differentiability of GC-GFRFT with Respect to Fractional Orders):}
Let $\mathbf{F}_{\mathrm{GC}}^{(\lambda;\alpha,\beta)}$ denote the GC-GFRFT operator defined in \eqref{eq:gc_vec}. For any fixed $\lambda$ satisfying Assumption~1, $\mathbf{F}_{\mathrm{GC}}^{(\lambda;\alpha,\beta)}$ is differentiable with respect to $\alpha$ and $\beta$. This differentiability enables the following risk-minimization formulation:
\begin{equation}
	\begin{aligned}
		\min_{\alpha,\beta,\,\mathbf{h}}
		\mathcal{L}(\alpha,\beta,\mathbf{h};\lambda) \\
		&\hspace{-7em}= \mathbb{E}\!\left\{ \left\|
		\big(\mathbf{F}_{\mathrm{GC}}^{(\lambda;\alpha,\beta)}\big)^{-1}
		\mathbf{H}_{\mathrm{GC}}
		\mathbf{F}_{\mathrm{GC}}^{(\lambda;\alpha,\beta)} \mathbf{y}
		-\mathbf{x}
		\right\|_2^2
		\right\}
	\end{aligned}
	\label{eq:obj-gc}
\end{equation}
where $\mathbf{H}_{\mathrm{GC}}=\mathrm{diag}(\mathbf{h})$ is the diagonal spectral filter in the GC-GFRFT domain. In implementation, the diagonal filtering corresponds to an element-wise multiplication in the matrix spectral domain, and \eqref{eq:obj-gc} is evaluated via \eqref{eq:gc_matrix} without explicitly forming $\mathbf{F}_{\mathrm{GC}}^{(\lambda;\alpha,\beta)}$.

The corresponding gradient-descent updates are given by
\begin{equation}
	\begin{bmatrix}
		\alpha^{(k+1)}\\
		\beta^{(k+1)}
	\end{bmatrix}
	=
	\begin{bmatrix}
		\alpha^{(k)}\\
		\beta^{(k)}
	\end{bmatrix}
	-\gamma
	\begin{bmatrix}
		\frac{\partial \mathcal{L}}{\partial \alpha}\\
		\frac{\partial \mathcal{L}}{\partial \beta}
	\end{bmatrix},
	\qquad
	\mathbf{h}^{(k+1)}=\mathbf{h}^{(k)}-\gamma_h\frac{\partial \mathcal{L}}{\partial \mathbf{h}}.
	\label{eq:update-rules-gc}
\end{equation}
Here, $\gamma$ and $\gamma_h$ denote the learning rates. Since the forward map in \eqref{eq:obj-gc} consists of differentiable matrix functions and linear operators, the gradients are well-defined.

\textit{Proof:} See Appendix~\ref{app:proof_thm2}.

The overall procedure is summarized in Algorithm~\ref{alg:gd-gc}.

\begin{algorithm}[htbp]
	\caption{GC-GFRFT Order and Filter Optimization}
	\label{alg:gd-gc}
	\begin{algorithmic}[1]
		\REQUIRE Observation $\mathbf{Y}$, fixed coupling parameter $\lambda$, learning rates $\gamma,\gamma_h$, iterations $L$
		\ENSURE Learned orders $(\alpha,\beta)$ and diagonal filter $\mathbf{H}_{\mathrm{GC}}$
		
		\STATE Initialize $\alpha=0.5$, $\beta=0.5$, and $\mathbf{H}_{\mathrm{GC}}=\mathbf{I}$
		\FOR{$k=1$ to $L$}
		\STATE Construct spatial basis $\mathbf{F}_{G_1}^{\alpha}$ and temporal coupling basis $\mathbf{F}_{t,\mathrm{GC}}^{(\lambda;\beta)}$ via \eqref{eq:gc_Ft}
		\STATE Compute the GC spectral representation:
		$\widehat{\mathbf{Y}} = \mathbf{F}_{G_1}^{\alpha} \mathbf{Y} \big(\mathbf{F}_{t,\mathrm{GC}}^{(\lambda;\beta)}\big)^{\mathrm{T}}$
		\STATE Apply element-wise spectral filtering:
		$\widehat{\mathbf{X}}_{\mathrm{spec}} = \mathbf{H}_{\mathrm{GC}}^{\mathrm{mat}} \circ \widehat{\mathbf{Y}}$
		\STATE Inverse transform:
		$\widehat{\mathbf{X}} = \big(\mathbf{F}_{G_1}^{\alpha}\big)^{\mathrm{H}} \widehat{\mathbf{X}}_{\mathrm{spec}} \big(\mathbf{F}_{t,\mathrm{GC}}^{(\lambda;\beta)}\big)^{*}$
		\STATE Evaluate the empirical loss corresponding to \eqref{eq:obj-gc}
		\STATE Update $(\alpha,\beta,\mathbf{h})$ according to \eqref{eq:update-rules-gc}
		\ENDFOR
		\RETURN $(\alpha,\beta)$ and $\mathbf{H}_{\mathrm{GC}}$
	\end{algorithmic}
\end{algorithm}

\subsection{Computational Efficiency and Optimization Analysis}
\label{subsec:efficiency_optimization}

This subsection analyzes the computational cost and optimization behavior of the proposed learnable Wiener filtering framework. 
The efficiency primarily stems from the separable implementation of the GC-GFRFT in \eqref{eq:gc_matrix}, which avoids explicitly forming the $(N_1N_2)\times(N_1N_2)$ Kronecker matrix. 
In addition, the phase-domain representation in \eqref{eq:gc_phase} clarifies how the fixed coupling parameter $\lambda$ enters the temporal operator through diagonal phase factors.

The overall computation consists of an offline spectral preprocessing stage and an online iterative learning stage. 
In the dense setting, preprocessing computes the eigendecompositions required by the spatial and temporal graph-induced fractional bases, incurring a one-time cost of $O(N_1^3+N_2^3)$ because the underlying graph topologies are fixed. 
These decompositions provide fixed eigenvector bases; during training, changing $\alpha$ or $\beta$ only updates the associated diagonal phase factors. 
Hence, the transform application reuses the cached eigenvectors and evaluates the operator via products of the form $\mathbf{V}\,\mathrm{diag}(\cdot)\,\mathbf{V}^{\mathrm H}$, \emph{without recomputing eigendecompositions across epochs}.

During learning, the forward and inverse GC-GFRFT evaluations are computed efficiently through sequential matrix multiplications in \eqref{eq:gc_matrix}. 
For $\mathbf{X}\in\mathbb{C}^{N_1\times N_2}$, each forward or inverse pass consists of one left multiplication by $\mathbf{F}_{G_1}^{\alpha}$ and one right multiplication by $\big(\mathbf{F}_{t,\mathrm{GC}}^{(\lambda;\beta)}\big)^{\mathrm T}$.
With dense arithmetic, each transform evaluation costs $O(N_1^2N_2+N_1N_2^2)$. 
The diagonal spectral filter $\mathbf{H}_{\mathrm{GC}}=\mathrm{diag}(\mathbf{h})$ is applied element-wise in the spectral domain and requires only $O(N_1N_2)$ operations. 
Consequently, the per-epoch complexity is dominated by a constant number of forward/inverse transform evaluations.

For the temporal coupling operator, \eqref{eq:gc_phase} shows that for a fixed $\beta$, the dependence on $\lambda$ is confined to diagonal phase factors. However, since $\beta$ is learnable, $\mathbf{W}_t(\beta)$ changes across epochs. Updating $\mathbf{F}_{t,\mathrm{GC}}^{(\lambda;\beta)}$ via \eqref{eq:gc_phase} requires recomputing a spectral factorization of the $N_2\times N_2$ unitary matrix $\mathbf{W}_t(\beta)$, which has a dense worst-case cost of $O(N_2^3)$. In typical spatiotemporal settings where $N_2\ll N_1$, this term is substantially smaller than the spatial transform cost $O(N_1^2N_2)$ and therefore does not change the overall scaling. Table~\ref{tab:comparison} summarizes the parameter counts and computational complexities of all considered transforms.
\begin{table}[htbp]
	\caption{Comparison of Learnable Parameters and Computational Complexity Among Different Fractional Transforms}
	\label{tab:comparison}
	\centering
	\resizebox{\columnwidth}{!}{
		\begin{tabular}{ccc} 
			\toprule
			Method & Learnable Parameters & Asymptotic Complexity \\
			\midrule
			2D-GFRFT   & $N_1N_2 + 1$ & $O(N_1^3 + N_2^3 + L(N_1^2N_2 + N_1N_2^2))$ \\
			2D-GBFRFT  & $N_1N_2 + 2$ & $O(N_1^3 + N_2^3 + L(N_1^2N_2 + N_1N_2^2))$ \\
			JFRFT      & $N_1N_2 + 2$ & $O(N_1^3 + N_2^3 + L(N_1^2N_2 + N_1N_2^2))$ \\
			GC-GFRFT   & $N_1N_2 + 2$ & $O(N_1^3 + N_2^3 + L(N_1^2N_2 + N_1N_2^2 + N_2^3))$ \\
			\bottomrule
		\end{tabular}
	}
	
	\vspace{4pt}
	\footnotesize
	\begin{tabular}{@{}p{\columnwidth}@{}}
		\textbf{Note:}
		$N_1$ and $N_2$ denote the sizes of the spatial and temporal graphs, and $L$ is the number of training epochs.
		The term $O(N_1^3+N_2^3)$ accounts for one-time offline eigendecompositions for graph-induced bases.
		For transforms involving a DFRFT temporal basis (e.g., JFRFT), the temporal basis can be precomputed once; we report a dense upper bound for consistency, while faster constructions may reduce this cost.
		The online cost is dominated by separable transform evaluations.
		For GC-GFRFT, when $\beta$ is learnable, updating $\mathbf{F}_{t,\mathrm{GC}}^{(\lambda;\beta)}$ requires recomputing a spectral factorization of $\mathbf{W}_t(\beta)$, introducing an additional $O(N_2^3)$ term per epoch.
		The coupling parameter $\lambda$ is treated as a fixed structural hyperparameter and is not counted as learnable.
	\end{tabular}
\end{table}

The learning objective exhibits a hybrid structure. For fixed $(\alpha,\beta,\lambda)$, optimizing the diagonal spectral filter coefficients $\mathbf{h}$ reduces to a convex quadratic problem. 
Optimizing the fractional orders $(\alpha,\beta)$ is generally non-convex because the GC-GFRFT operators depend nonlinearly on these parameters; accordingly, first-order methods converge to stationary points in general. 
In our experiments, the proposed operator family is differentiable with respect to $(\alpha,\beta)$, and standard optimizers exhibit stable convergence under the initialization described in Algorithm~\ref{alg:gd-gc}.

\begin{table*}[!t]
	\centering

	\setlength{\tabcolsep}{1.5pt}   
	\renewcommand{\arraystretch}{0.9} 
	
	\providecommand{\mygroupheader}[1]{
		\midrule
		\multicolumn{10}{c}{\textbf{#1}}\\[-0.4ex]
		\midrule
	}
	

	\caption{DENOISING RESULTS (MSE) ON THE COVID-19 DATASET}
	\label{tab:COVID_denoising_results_combined}
	\resizebox{\textwidth}{!}{
		\scriptsize
		\begin{tabular}{@{}c*{9}{c}@{}}
			\toprule
			\multirow{4}{*}{Method} & \multicolumn{3}{c}{3-NN} & \multicolumn{3}{c}{4-NN} & \multicolumn{3}{c}{5-NN} \\
			\cmidrule(lr){2-4} \cmidrule(lr){5-7} \cmidrule(lr){8-10}
			& $\sigma$=0.6 & $\sigma$=0.9 & $\sigma$=1.2
			& $\sigma$=0.6 & $\sigma$=0.9 & $\sigma$=1.2
			& $\sigma$=0.6 & $\sigma$=0.9 & $\sigma$=1.2 \\
			
			\mygroupheader{Fractional Transform-Based Methods}
			
			\makecell{2D-GFRFT \\ $(\alpha)$}
			& \makecell{0.6424 \\ \scriptsize (0.307)}
			& \makecell{1.0626 \\ \scriptsize (=0.263)}
			& \makecell{1.0749 \\ \scriptsize (-0.485)}
			& \makecell{0.4901 \\ \scriptsize (0.285)}
			& \makecell{0.8104 \\ \scriptsize (0.298)}
			& \makecell{1.1075 \\ \scriptsize (0.788)}
			& \makecell{0.3828 \\ \scriptsize (0.682)}
			& \makecell{0.5896 \\ \scriptsize (0.860)}
			& \makecell{0.7687 \\ \scriptsize (1.184)} \\
			
			\makecell{2D-GBFRFT \\ $(\alpha_1, \alpha_2)$}
			& \makecell{0.1060 \\ (0.033, 0.999)} & \makecell{\textbf{0.2155} \\ (0.047, 0.998)} & \makecell{\textbf{0.3459} \\ (0.060, 0.998)}
			& \makecell{0.4012 \\ (1.358, 0.594)} & \makecell{0.6977 \\ (1.362, 0.566)} & \makecell{1.1212 \\ (2.129, 0.565)}
			& \makecell{\textbf{0.1077} \\ (0.008, 1.000)} & \makecell{\textbf{0.2216} \\ (0.005, 1.000)} & \makecell{\textbf{0.3593} \\ (-0.002, 1.000)} \\
			
			\makecell{JFRFT \\ $(\alpha, \beta)$}
			& \makecell{\textbf{0.1051} \\ (0.051, 0.996)} & \makecell{0.2164 \\ (0.072, 0.995)} & \makecell{0.7659 \\ (1.055, 1.082)}
			& \makecell{0.4095 \\ (1.281, 0.609)} & \makecell{0.7371 \\ (1.274, 0.611)} & \makecell{\textbf{0.3538} \\ (0.057, 1.000)}
			& \makecell{0.7058 \\ (0.547, 0.318)} & \makecell{1.1149 \\ (0.553, 0.281)} & \makecell{0.3693 \\ (-0.040, 1.000)} \\
			
			\makecell{GC-GFRFT \\ $(\alpha,\beta,\lambda)$}
			& \makecell{\textbf{0.1051} \\ (0.051, 0.996, 1.000)} & \makecell{\textbf{0.2155} \\ (0.047, 0.998, 0.000)} & \makecell{\textbf{0.3459} \\ (0.060, 0.998, 0.000)}
			& \makecell{\textbf{0.1697} \\ (0.060, 0.715, 0.800)} & \makecell{\textbf{0.2740} \\ (0.071, 1.019, 0.800)} & \makecell{\textbf{0.3538} \\ (0.057, 1.000, 1.000)}
			& \makecell{\textbf{0.1077} \\ (0.008, 1.000, 0.000)} & \makecell{\textbf{0.2216} \\ (0.005, 1.000, 0.000)} & \makecell{\textbf{0.3593} \\ (-0.002, 1.000, 0.000)} \\
			
			\mygroupheader{Traditional Baselines}
			
			SGWT   & 3.5500 & 6.8600 & 11.5600 & 3.5200 & 6.8400 & 11.5600 & 3.5000 & 6.7900 & 11.5100 \\
			Median & 4.5300 & 9.5900 & 17.2800 & 5.2500 & 9.9700 & 16.5400 & 4.5700 & 9.0500 & 14.9300 \\
			ARMA   & 2.4300 & 3.3200 & 4.2100  & 2.4300 & 2.9800 & 3.4900  & 2.3100 & 2.7200 & 3.1000  \\
			
			\mygroupheader{Graph Neural Network Baselines}
			
			GCN & 1.2613 & 1.3116 & 1.3455 & 1.3972 & 1.4315 & 1.4547 & 1.4741 & 1.5060 & 1.5163 \\
			GAT & 1.3306 & 1.2610 & 1.2498 & 1.3729 & 1.4338 & 1.5736 & 1.4473 & 1.4668 & 1.5487 \\
			ChebyNet & 0.9834 & 1.1955 & 1.3489 & 0.7331 & 0.9758 & 1.1557 & 0.4505 & 0.5852 & 0.6853 \\
			
			\bottomrule
		\end{tabular}
	}

	\vspace{0.3cm}

	\caption{DENOISING RESULTS (MSE) ON THE SST DATASET}
	\label{tab:SST_denoising_results_combined}
	\resizebox{\textwidth}{!}{
		\scriptsize
		\begin{tabular}{@{}c*{9}{c}@{}}
			\toprule
			\multirow{4}{*}{Method} & \multicolumn{3}{c}{3-NN} & \multicolumn{3}{c}{4-NN} & \multicolumn{3}{c}{5-NN} \\
			\cmidrule(lr){2-4} \cmidrule(lr){5-7} \cmidrule(lr){8-10}
			& $\sigma$=0.6 & $\sigma$=0.9 & $\sigma$=1.2
			& $\sigma$=0.6 & $\sigma$=0.9 & $\sigma$=1.2
			& $\sigma$=0.6 & $\sigma$=0.9 & $\sigma$=1.2 \\
			
			\mygroupheader{Fractional Transform-Based Methods}
			
			\makecell{2D-GFRFT \\ $(\alpha)$}
			& \makecell{0.7200 \\ \scriptsize (0.698)}
			& \makecell{1.5089 \\ \scriptsize (0.684)}
			& \makecell{2.4722 \\ \scriptsize (0.668)}
			& \makecell{0.7888 \\ \scriptsize (0.262)}
			& \makecell{1.6653 \\ \scriptsize (0.262)}
			& \makecell{2.7415 \\ \scriptsize (0.265)}
			& \makecell{0.6310 \\ \scriptsize (0.783)}
			& \makecell{1.2809 \\ \scriptsize (0.792)}
			& \makecell{2.0307 \\ \scriptsize (0.820)} \\
			
			\makecell{2D-GBFRFT \\ $(\alpha_1, \alpha_2)$}
			& \makecell{0.6534 \\ (0.582, 0.740)}
			& \makecell{\textbf{1.3608} \\ (0.561, 0.755)}
			& \makecell{\textbf{2.2075} \\ (0.522, 0.774)}
			& \makecell{0.5552 \\ (0.181, 0.827)}
			& \makecell{\textbf{1.0648} \\ (0.163, 1.003)}
			& \makecell{1.7662 \\ (0.165, 1.000)}
			& \makecell{0.6064 \\ (0.739, 0.379)}
			& \makecell{1.2161 \\ (0.753, 0.484)}
			& \makecell{1.9799 \\ (0.762, 0.517)} \\
			
			\makecell{JFRFT \\ $(\alpha, \beta)$}
			& \makecell{1.1685 \\ (0.752, 0.747)}
			& \makecell{2.2851 \\ (0.748, 0.779)}
			& \makecell{3.4158 \\ (0.850, 0.889)}
			& \makecell{2.1764 \\ (0.430, 0.447)}
			& \makecell{1.1908 \\ (0.185, 0.994)}
			& \makecell{2.0045 \\ (0.185, 1.000)}
			& \makecell{\textbf{0.4772} \\ (0.740, 0.642)}
			& \makecell{\textbf{0.8394} \\ (0.667, 0.759)}
			& \makecell{\textbf{1.4031} \\ (0.669, 0.744)} \\
			
			\makecell{GC-GFRFT \\ $(\alpha,\beta,\lambda)$}
			& \makecell{\textbf{0.6395} \\ (0.604, 0.717, 0.100)}
			& \makecell{\textbf{1.3608} \\ (0.561, 0.755, 0.000)}
			& \makecell{\textbf{2.2075} \\ (0.522, 0.774, 0.000)}
			& \makecell{\textbf{0.5295} \\ (0.238, 1.238, 0.400)}
			& \makecell{\textbf{1.0648} \\ (0.163, 1.003, 0.000)}
			& \makecell{\textbf{1.2807} \\ (1.098, 1.289, 0.600)}
			& \makecell{\textbf{0.4772} \\ (0.740, 0.642, 1.000)}
			& \makecell{\textbf{0.8394} \\ (0.667, 0.759, 1.000)}
			& \makecell{\textbf{1.4031} \\ (0.669, 0.744, 1.000)} \\
			
			\mygroupheader{Traditional Baselines}
			
			SGWT   & 5.3200 & 10.7100 & 18.2900 & 5.3000 & 10.6200 & 18.1200 & 4.8400 & 9.7700 & 16.7800 \\
			Median & 6.0400 & 13.7600 & 24.2800 & 6.3400 & 12.6900 & 21.6500 & 5.4000 & 10.0900 & 16.8300 \\
			ARMA   & 2.9700 & 4.3900  & 5.7900  & 3.2400 & 4.3500  & 5.2800  & 3.7300 & 4.7800  & 5.6300  \\
			
			\mygroupheader{Graph Neural Network Baselines}
			
			GCN & 2.5521 & 3.6928 & 4.4128 & 1.9366 & 2.6767 & 3.3740 & 2.1656 & 2.9620 & 3.7989 \\
			GAT & 2.5520 & 3.7652 & 4.9559 & 3.0085 & 4.2111 & 5.4557 & 3.2268 & 4.5049 & 5.9509 \\
			ChebyNet & 3.4350 & 4.7864 & 5.4158 & 3.0030 & 4.3172 & 5.3172 & 2.6764 & 4.0671 & 4.8447 \\
			
			\bottomrule
		\end{tabular}
	}

	\vspace{0.3cm}

	\caption{DENOISING RESULTS (MSE) ON THE PM-25 DATASET}
	\label{tab:PM25_denoising_results_combined}
	\resizebox{\textwidth}{!}{
		\scriptsize
		\begin{tabular}{@{}c*{9}{c}@{}}
			\toprule
			\multirow{4}{*}{Method} & \multicolumn{3}{c}{3-NN} & \multicolumn{3}{c}{4-NN} & \multicolumn{3}{c}{5-NN} \\
			\cmidrule(lr){2-4} \cmidrule(lr){5-7} \cmidrule(lr){8-10}
			& $\sigma$=0.6 & $\sigma$=0.9 & $\sigma$=1.2
			& $\sigma$=0.6 & $\sigma$=0.9 & $\sigma$=1.2
			& $\sigma$=0.6 & $\sigma$=0.9 & $\sigma$=1.2 \\
			
			\mygroupheader{Fractional Transform-Based Methods}
			
			\makecell{2D-GFRFT \\ $(\alpha)$}
			& \makecell{0.6636 \\ \scriptsize (0.813)}
			& \makecell{1.2474 \\ \scriptsize (0.804)}
			& \makecell{1.8847 \\ \scriptsize (0.801)}
			& \makecell{0.9222 \\ \scriptsize (0.584)}
			& \makecell{1.7594 \\ \scriptsize (0.577)}
			& \makecell{2.1018 \\ \scriptsize (1.014)}
			& \makecell{0.6605 \\ \scriptsize (0.315)}
			& \makecell{1.1877 \\ \scriptsize (0.318)}
			& \makecell{1.7555 \\ \scriptsize (0.320)} \\
			
			\makecell{2D-GBFRFT \\ $(\alpha_1, \alpha_2)$}
			& \makecell{0.6398 \\ (0.784, 1.371)} & \makecell{1.1952 \\ (0.723, 0.513)} & \makecell{1.8206 \\ (0.716, 0.507)}
			& \makecell{0.7653 \\ (0.613, -0.273)} & \makecell{1.5059 \\ (0.618, 0.272)} & \makecell{2.2999 \\ (0.624, 0.275)}
			& \makecell{0.6008 \\ (0.474, 0.527)} & \makecell{1.1854 \\ (0.322, 0.298)} & \makecell{1.7549 \\ (0.322, 0.311)} \\
			
			\makecell{JFRFT \\ $(\alpha, \beta)$}
			& \makecell{0.4872 \\ (0.785, 0.517)} & \makecell{0.8825 \\ (0.846, 0.573)} & \makecell{1.3697 \\ (0.827, 0.563)}
			& \makecell{\textbf{0.4219} \\ (0.965, 0.561)} & \makecell{\textbf{0.8629} \\ (0.977, 0.567)} & \makecell{1.3931 \\ (0.974, 0.564)}
			& \makecell{0.4497 \\ (0.480, 0.541)} & \makecell{0.9242 \\ (0.490, 0.538)} & \makecell{1.4784 \\ (0.505, 0.536)} \\
			
			\makecell{GC-GFRFT \\ $(\alpha,\beta,\lambda)$}
			& \makecell{\textbf{0.3746} \\ (0.857, 0.868, 0.700)} & \makecell{\textbf{0.7509} \\ (0.852, 0.855, 0.700)} & \makecell{\textbf{1.1665} \\ (0.701, 1.398, 0.300)}
			& \makecell{\textbf{0.4219} \\ (0.965, 0.561, 1.000)} & \makecell{\textbf{0.8629} \\ (0.977, 0.567, 1.000)} & \makecell{\textbf{1.3445} \\ (1.025, 0.944, 0.600)}
			& \makecell{\textbf{0.4191} \\ (0.418, 0.656, 0.600)} & \makecell{\textbf{0.8275} \\ (1.078, 1.475, 0.300)} & \makecell{\textbf{1.3906} \\ (0.414, 0.686, 0.600)} \\
			
			\mygroupheader{Traditional Baselines}
			
			SGWT   & 3.6100 & 6.6500 & 10.9800 & 3.3700 & 6.1900 & 10.2300 & 3.3700 & 6.2000 & 10.2400 \\
			Median & 5.0300 & 8.9600 & 14.7000 & 4.3800 & 7.4200 & 11.8700 & 4.6700 & 7.7800 & 12.0500 \\
			ARMA   & 2.6200 & 3.3000 & 3.9700  & 2.5000 & 3.0300 & 3.5900  & 2.4000 & 2.7500 & 3.0800  \\
			
			\mygroupheader{Graph Neural Network Baselines}
			
			GCN & 0.7009 & 0.7143 & 0.6977 & 0.7127 & 0.7210 & 0.7150 & 0.7032 & 0.7084 & 0.6992 \\
			GAT & 0.8098 & 0.7671 & 0.7785 & 0.8110 & 0.7697 & 0.7958 & 0.7495 & 0.7194 & 0.7359 \\
			ChebyNet & 0.6436 & 0.6625 & 0.6766 & 0.5952 & 0.6204 & 0.6527 & 0.6026 & 0.6215 & 0.6459 \\
			
			\bottomrule
		\end{tabular}
	} 
\end{table*}

\section{Applications on Real-World Data}
\label{sec:realdata}

\subsection{Denoising of Time-Varying Graph Signals}
In~\cite{1}, time-varying graph signals are modeled by representing the temporal dimension as a path graph and the spatial relationships by another graph, thereby forming a Cartesian product graph. This spatiotemporal modeling protocol enables a direct comparison with JFRFT under the same data and graph-construction setting. To evaluate denoising performance, we apply our framework to three real-world spatiotemporal datasets adopted from the benchmark setting in~\cite{40}, namely COVID-19, Sea surface temperature (SST), and PM-25 air quality. For each dataset, the spatial relationships are modeled by $k$-nearest neighbor graphs, abbreviated as $k$-NN, with $k\in\{3,4,5\}$, and corrupted observations are generated by injecting additive white Gaussian noise (AWGN) with standard deviation $\sigma\in\{0.6,0.9,1.2\}$ into the clean spatiotemporal signals.

To ensure a principled comparison within the class of fractional transform-based approaches, we employ the same GD optimization protocol for 2D-GFRFT, 2D-GBFRFT, JFRFT, and the proposed GC-GFRFT. Specifically, the learning rate is set to $0.1$ and the number of iterations is fixed to $200$. All fractional orders are initialized to $0.5$, and the diagonal spectral filter is initialized as the identity matrix. This unified setup isolates the effect of the underlying fractional spectral representations from confounding factors arising from heterogeneous optimization choices.

For GC-GFRFT, the coupling regularization parameter $\lambda$ is selected via an outer grid search with step size $0.1$ and then kept fixed during optimization. Thus, $\lambda$ is excluded from the learnable-parameter count and is not updated by gradient-based learning. For each candidate $\lambda$, the remaining fractional orders and the diagonal filter are optimized using the same GD protocol described above, and the reported results correspond to the best-performing $\lambda$ under the adopted validation criterion.

As reported in Tables~\ref{tab:COVID_denoising_results_combined}--\ref{tab:PM25_denoising_results_combined}, 2D-GBFRFT improves upon 2D-GFRFT in most settings, which supports the advantage of decoupling fractional orders across spatial and temporal dimensions. A small number of configurations exhibit inferior performance, which is consistent with the nonconvexity induced by fractional spectral parameterization and the sensitivity of first-order optimization to initialization and local stationary points. Moreover, the relative ordering between 2D-GBFRFT and JFRFT varies across datasets and graph constructions, indicating that no single fixed fractional basis is uniformly optimal for real-world time-varying graph signals.

Relative to classical filtering baselines, including the spectral graph wavelet transform (SGWT)~\cite{16}, median filtering~\cite{58}, and autoregressive moving-average (ARMA) graph filtering~\cite{22}, fractional transform-based methods achieve substantially lower MSE across nearly all noise levels and $k$-NN settings. These results suggest that adaptive fractional spectral representations provide a stronger denoising prior than hand-crafted smoothing operators.

Finally, GC-GFRFT attains the best or tied-best performance in the majority of configurations, demonstrating robust effectiveness across heterogeneous spatiotemporal characteristics. We further compare GC-GFRFT with representative graph neural network baselines, including the graph convolutional network (GCN)~\cite{59}, the graph attention network (GAT)~\cite{60}, and Chebyshev graph convolution (ChebyNet)~\cite{61}. Since neural models are commonly trained under model-specific practices such as learning-rate schedules, numbers of epochs, and early-stopping criteria, these baselines are reported under their standard training and model-selection protocols rather than being constrained to the GD setup adopted for fractional transforms. Under these typical settings, GC-GFRFT remains highly competitive and often yields lower MSE; however, in a limited number of configurations, a neural baseline can achieve slightly better performance. Overall, the results demonstrate that GC-GFRFT consistently surpasses classical filtering baselines while providing performance that is superior or comparable to representative neural approaches on real-world time-varying graph signals.

\subsection{Deblurring of Dynamic Images}
We further evaluate the proposed GC-GFRFT on dynamic image deblurring and compare it with fractional transform baselines, including 2D-GFRFT, 2D-GBFRFT, and JFRFT. In addition, we report results of three representative learning-based deblurring methods, DeepDeblur~\cite{62}, Restormer~\cite{63}, and NAFNet~\cite{64}, as external baselines. Experiments are conducted on the REDS dataset. We extract three consecutive frames, resize them to $200\times200$ pixels, and partition each frame into $100$ non-overlapping patches of size $20\times20$. Each triplet of corresponding patches across the three frames is modeled as a spatiotemporal graph signal, where grayscale intensities are treated as signal values. The spatial graph is constructed as a $4$-nearest neighbor graph based on pixel locations, and the temporal dimension is represented by a $3$-node path graph.

Each $20\times20$ patch triplet is treated as an independent instance for optimization. The fractional orders and diagonal filter coefficients are optimized in a patch-wise manner, and the restored patches are reassembled to reconstruct the deblurred sequence. Adaptive filtering is performed under the respective transforms (2D-GFRFT, 2D-GBFRFT, JFRFT, and GC-GFRFT), with MSE used as the optimization objective. Optimization is carried out using Adam with a learning rate of $2\times10^{-2}$ for $100$ epochs. The fractional orders are initialized to $0.8$, and the diagonal filter is initialized as the identity matrix.

For GC-GFRFT, the coupling parameter $\lambda$ is selected by an outer grid search with step size $0.1$ and is kept fixed during the subsequent optimization. Gradient-based updates are applied only to the fractional orders and the diagonal spectral filter coefficients; in particular, $\lambda$ is not updated by backpropagation and is not counted as a learnable parameter.

For each frame $t$, we compute MSE, peak signal-to-noise ratio (PSNR), and structural similarity index measure (SSIM) as follows:
\begin{equation}
	MSE_t = \frac{1}{HW}\sum_{i,j}\left(I_t(i,j)-\hat{I}_t(i,j)\right)^2,
\end{equation}
\begin{equation}
	PSNR_t = 10\log_{10}\!\left(\frac{\mathrm{MAX}^2}{MSE_t}\right),
\end{equation}
\begin{equation}
	SSIM_t = \frac{(2\mu_{I_t}\mu_{\hat{I}_t}+C_1)(2\sigma_{I_t\hat{I}_t}+C_2)}
	{(\mu_{I_t}^2+\mu_{\hat{I}_t}^2+C_1)(\sigma_{I_t}^2+\sigma_{\hat{I}_t}^2+C_2)},
\end{equation}
where $I_t$ and $\hat{I}_t$ denote the ground-truth and reconstructed frames at time $t$, respectively, and $H$ and $W$ are the image height and width.

Quantitative results are reported in Tables~\ref{tab:deblurring_results_single} and~\ref{tab:deblurring_results_avg}. Among the fractional transform baselines, 2D-GBFRFT consistently improves upon 2D-GFRFT, supporting the benefit of allowing distinct fractional orders along the spatial and temporal dimensions. GC-GFRFT further reduces $MSE_{\mathrm{avg}}$ and achieves the best overall performance in terms of $MSE_{\mathrm{avg}}$, $PSNR_{\mathrm{avg}}$, and $SSIM_{\mathrm{avg}}$, indicating that coupling-aware fractional modeling provides a more effective transform-domain representation for dynamic deblurring under patch-wise optimization.

For the fractional transform-based methods, we adopt an identical patch-wise instance-adaptive optimization protocol, where the fractional orders and diagonal spectral filter are updated by Adam for each patch triplet. In contrast, the learning-based baselines are evaluated in inference mode using their official implementations and publicly released weights, without any test-time parameter adaptation. Therefore, comparisons with learning-based methods primarily reflect the effectiveness of instance-adaptive fractional spectral modeling under short temporal windows, rather than a fully matched training paradigm.

\begin{figure*}[t!]
\centering
\setlength{\tabcolsep}{1.2pt}
\renewcommand{\arraystretch}{0}
\newcommand{\imgw}{0.107\textwidth}

\begin{tabular}{ccccccccc}
\includegraphics[width=\imgw]{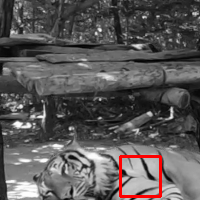} &
\includegraphics[width=\imgw]{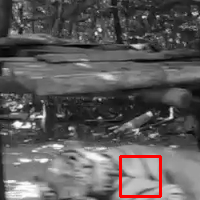} &
\includegraphics[width=\imgw]{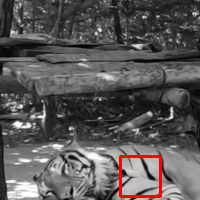} &
\includegraphics[width=\imgw]{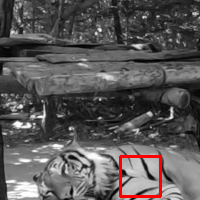} &
\includegraphics[width=\imgw]{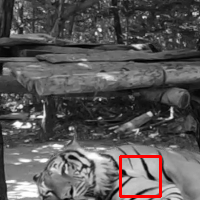} &
\includegraphics[width=\imgw]{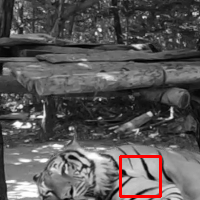} &
\includegraphics[width=\imgw]{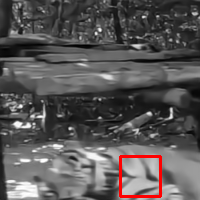} &
\includegraphics[width=\imgw]{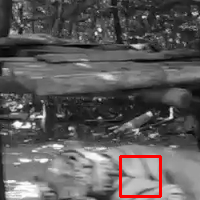} &
\includegraphics[width=\imgw]{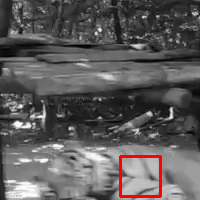} \\[0.6ex]

\includegraphics[width=\imgw]{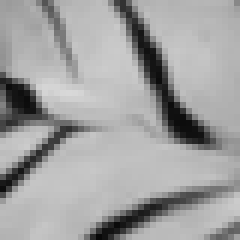} &
\includegraphics[width=\imgw]{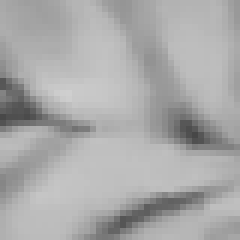} &
\includegraphics[width=\imgw]{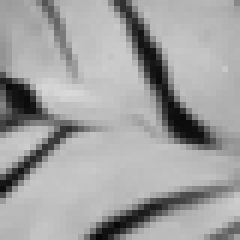} &
\includegraphics[width=\imgw]{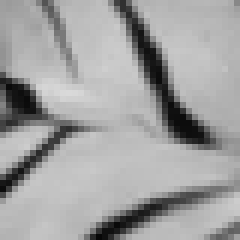} &
\includegraphics[width=\imgw]{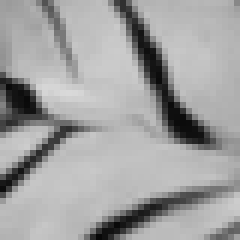} &
\includegraphics[width=\imgw]{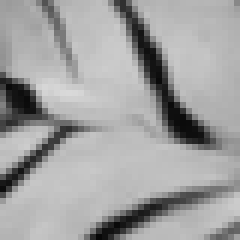} &
\includegraphics[width=\imgw]{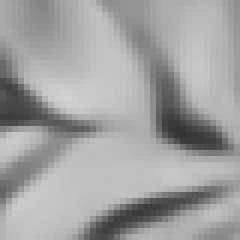} &
\includegraphics[width=\imgw]{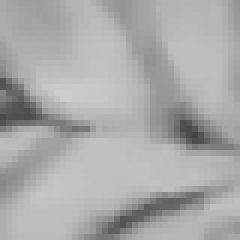} &
\includegraphics[width=\imgw]{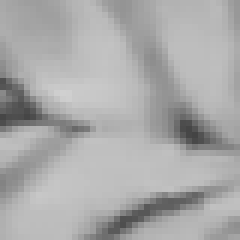} \\[1.2ex]

\includegraphics[width=\imgw]{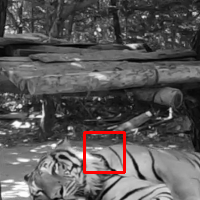} &
\includegraphics[width=\imgw]{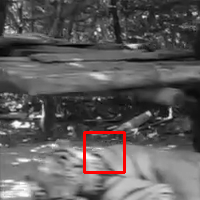} &
\includegraphics[width=\imgw]{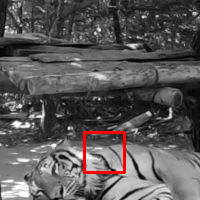} &
\includegraphics[width=\imgw]{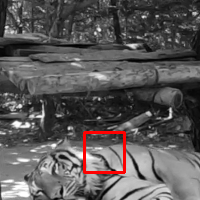} &
\includegraphics[width=\imgw]{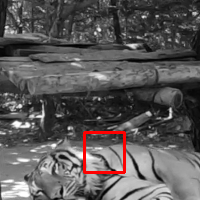} &
\includegraphics[width=\imgw]{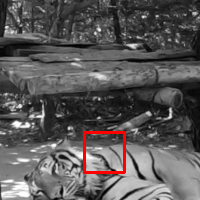} &
\includegraphics[width=\imgw]{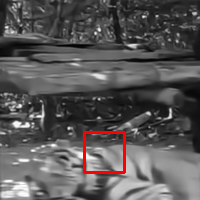} &
\includegraphics[width=\imgw]{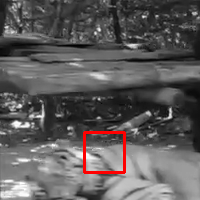} &
\includegraphics[width=\imgw]{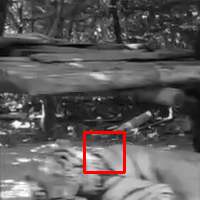} \\[0.6ex]

\includegraphics[width=\imgw]{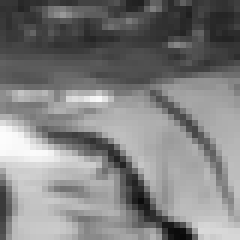} &
\includegraphics[width=\imgw]{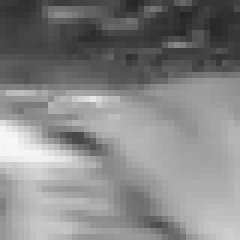} &
\includegraphics[width=\imgw]{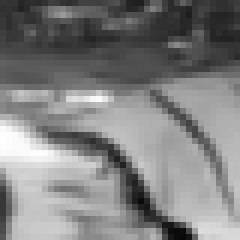} &
\includegraphics[width=\imgw]{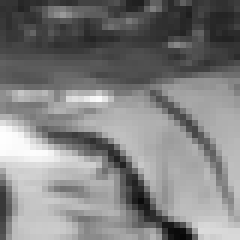} &
\includegraphics[width=\imgw]{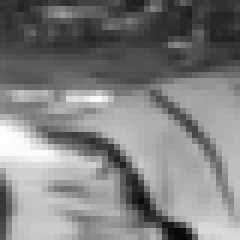} &
\includegraphics[width=\imgw]{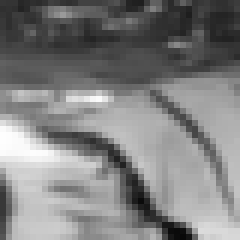} &
\includegraphics[width=\imgw]{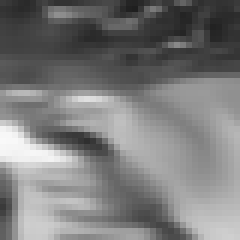} &
\includegraphics[width=\imgw]{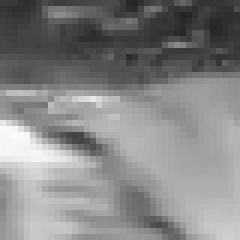} &
\includegraphics[width=\imgw]{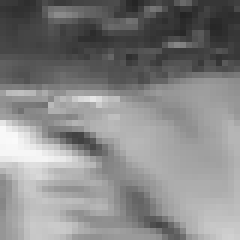} \\[1.2ex]

\includegraphics[width=\imgw]{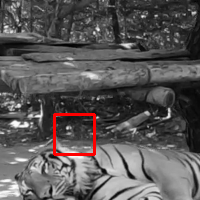} &
\includegraphics[width=\imgw]{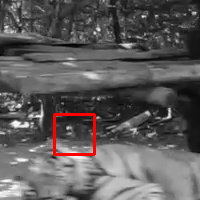} &
\includegraphics[width=\imgw]{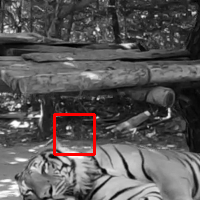} &
\includegraphics[width=\imgw]{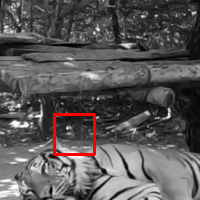} &
\includegraphics[width=\imgw]{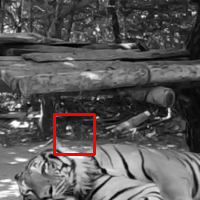} &
\includegraphics[width=\imgw]{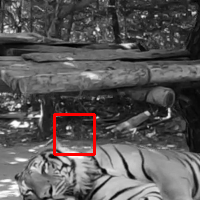} &
\includegraphics[width=\imgw]{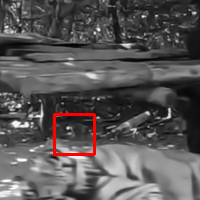} &
\includegraphics[width=\imgw]{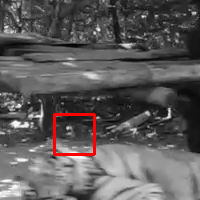} &
\includegraphics[width=\imgw]{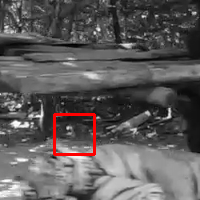} \\[0.6ex]

\includegraphics[width=\imgw]{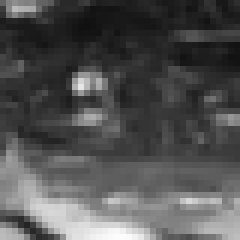} &
\includegraphics[width=\imgw]{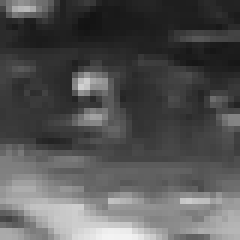} &
\includegraphics[width=\imgw]{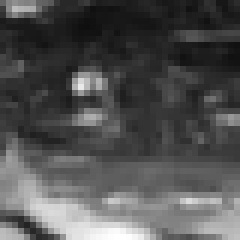} &
\includegraphics[width=\imgw]{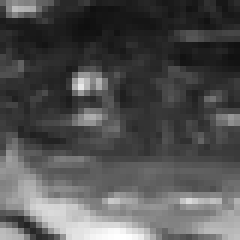} &
\includegraphics[width=\imgw]{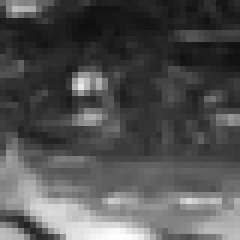} &
\includegraphics[width=\imgw]{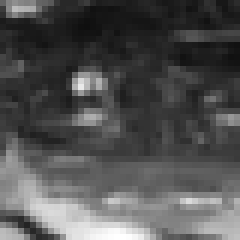} &
\includegraphics[width=\imgw]{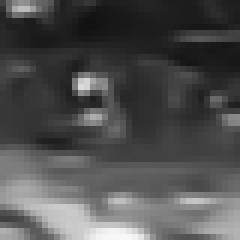} &
\includegraphics[width=\imgw]{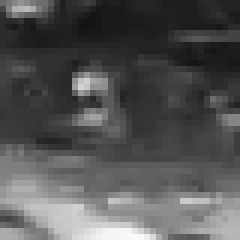} &
\includegraphics[width=\imgw]{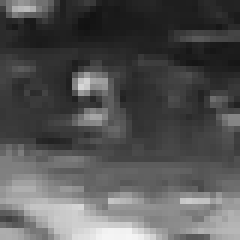} \\[1.2ex]

\scriptsize Original & \scriptsize Blurred & \scriptsize 2D-GFRFT & \scriptsize 2D-GBFRFT & \scriptsize JFRFT & \scriptsize GC-GFRFT & \scriptsize DeepDeblur & \scriptsize Restormer & \scriptsize NAFNet
\end{tabular}

\caption{Visual comparison on the REDS sequence for dynamic image deblurring. For each frame, the first row shows the selected region on the full, and the second row shows the corresponding magnified patch.}
\label{fig:reds_zoom_compare1}
\end{figure*}

Fig.~\ref{fig:reds_zoom_compare1} provides qualitative comparisons for dynamic image deblurring. For each frame, a representative region is highlighted on the full image, and the corresponding magnified patch is shown beneath it. Compared with the competing methods, GC-GFRFT recovers sharper edges and clearer local textures, while exhibiting fewer residual blur artifacts around object boundaries and fine structures. These visual observations are consistent with the quantitative results and further demonstrate the effectiveness of GC-GFRFT for dynamic image deblurring.
\begin{table}[htbp]
	\centering
	\caption{Comparison of Deblurring Performance (MSE/PSNR/SSIM) on the REDS Dataset}
	\label{tab:deblurring_results_single}
	\resizebox{\columnwidth}{!}
	{
		\begin{tabular}{cccccc}
			\toprule
			Dataset & Method & Frame & MSE & PSNR & SSIM \\
			\midrule
			
			\multirow{21}{*}[-4.2ex]{REDS}
			& \multirow{3}{*}{2D-GFRFT}
			& Frame 1 & 1.0389 & 47.96 & 0.9992 \\
			& & Frame 2 & 0.9281 & 48.45 & 0.9992 \\
			& & Frame 3 & 1.2405 & 47.20 & 0.9991 \\
			\cmidrule(lr){2-6}
			
			& \multirow{3}{*}{2D-GBFRFT}
			& Frame 1 & 1.0010 & 48.13 & 0.9993 \\
			& & Frame 2 & \textbf{0.9137} & \textbf{48.52} & \textbf{0.9993} \\
			& & Frame 3 & 1.2424 & 47.19 & 0.9992 \\
			\cmidrule(lr){2-6}
			
			& \multirow{3}{*}{JFRFT}
			& Frame 1 & \textbf{0.9406} & \textbf{48.40} & \textbf{0.9994} \\
			& & Frame 2 & 0.9392 & 48.40 & 0.9993 \\
			& & Frame 3 & 1.2460 & 47.18 & 0.9991 \\
			\cmidrule(lr){2-6}
			
			& \multirow{3}{*}{GC-GFRFT}
			& Frame 1 & 0.9564 & 48.32 & 0.9994 \\
			& & Frame 2 & 0.9308 & 48.44 & 0.9993 \\
			& & Frame 3 & \textbf{1.2107} & \textbf{47.30} & \textbf{0.9992} \\
			\cmidrule(lr){2-6}
			
			& \multirow{3}{*}{DeepDeblur}
			& Frame 1 & 204.6643 & 25.02 & 0.8140 \\
			& & Frame 2 & 214.1956 & 24.82 & 0.8054 \\
			& & Frame 3 & 229.0492 & 24.53 & 0.8028 \\
			\cmidrule(lr){2-6}
			
			& \multirow{3}{*}{Restormer}
			& Frame 1 & 253.9732 & 24.08 & 0.7949 \\
			& & Frame 2 & 245.9362 & 24.22 & 0.7961 \\
			& & Frame 3 & 262.5448 & 23.94 & 0.7934 \\
			\cmidrule(lr){2-6}
			
			& \multirow{3}{*}{NAFNet}
			& Frame 1 & 245.4231 & 24.23 & 0.7994 \\
			& & Frame 2 & 238.4010 & 24.36 & 0.7997 \\
			& & Frame 3 & 266.5690 & 23.87 & 0.7901 \\
			\bottomrule
	\end{tabular}}
\end{table}
\begin{table}[htbp]
\centering
\caption{Deblurring Performance in Terms of $MSE_{\text{avg}}$, $PSNR_{\text{avg}}$, and $SSIM_{\text{avg}}$ on the REDS Dataset}
\label{tab:deblurring_results_avg}
\resizebox{\columnwidth}{!}
{
\begin{tabular}{ccccc}
\toprule
Dataset & Method & $MSE_{\text{avg}}$ & $PSNR_{\text{avg}}$ & $SSIM_{\text{avg}}$ \\
\midrule
\multirow{7}{*}{REDS}
& 2D-GFRFT   & 1.0692 & 47.87 & 0.9992 \\
& 2D-GBFRFT  & 1.0524 & 47.95 & 0.9992 \\
& JFRFT      & 1.0419 & 47.99 & 0.9993 \\
& GC-GFRFT   & \textbf{1.0326} & \textbf{48.02} & \textbf{0.9993} \\
& DeepDeblur & 215.9697 & 24.79 & 0.8074 \\
& Restormer  & 254.1514 & 24.08 & 0.7948 \\
& NAFNet     & 250.1310 & 24.15 & 0.7964 \\
\bottomrule
\end{tabular}}
\end{table}

\subsection{Denoising of Dynamic Images}

To further evaluate robustness beyond dynamic deblurring, we consider a dynamic image denoising task on the REDS dataset. Clean sequences are first converted to floating-point representations and normalized to the range $[0,1]$. Additive white Gaussian noise (AWGN) with zero mean and standard deviation $\sigma/255$ is then injected into the normalized intensities, where we set $\sigma=18$ on the 8-bit scale. After noise injection, the corrupted frames are clipped to the valid range $[0,1]$ and re-quantized to 8-bit integers, which follows a standard digital imaging pipeline and ensures valid pixel representations.

We compare the proposed GC-GFRFT with fractional transform baselines, including 2D-GFRFT, 2D-GBFRFT, and JFRFT. We also include three classical spatial denoisers as reference baselines, namely Gaussian filtering, median filtering, and non-local means (NLM), following the implementation and parameter settings in~\cite{41}. These classical methods are applied frame-wise without any learning or test-time adaptation. Gaussian filtering uses a $5\times5$ kernel with the standard deviation determined by the kernel size. Median filtering uses a $5\times5$ neighborhood. NLM denoising uses filter strength $h=20$ with a $7\times7$ template window and a $21\times21$ search window. For the fractional transform-based methods, all experimental settings---including the patch-wise spatiotemporal graph construction, initialization strategy, optimization hyperparameters, and stopping criterion---are kept identical to those used in the preceding dynamic deblurring experiments unless otherwise specified, thereby ensuring a consistent and fair comparison within the fractional transform family.

Quantitative results are reported in Tables~\ref{tab:denoising_results_single} and~\ref{tab:denoising_results_avg}. Overall, the fractional transform-based methods consistently outperform classical spatial filters across the evaluated frames. This indicates that exploiting spatiotemporal structure through graph-based spectral representations provides a stronger denoising prior than purely spatial neighborhood smoothing. Within the fractional transform family, decoupling the fractional orders along the spatial and temporal dimensions generally improves over using a single shared order, consistent with the benefit of increased modeling flexibility. The proposed GC-GFRFT achieves the best overall average performance and remains highly competitive at the per-frame level, supporting the effectiveness of coupling-aware fractional modeling for dynamic denoising.

\begin{figure*}[t!]
\centering
\setlength{\tabcolsep}{1.2pt}
\renewcommand{\arraystretch}{0}
\newcommand{\imgw}{0.107\textwidth}

\begin{tabular}{ccccccccc}
\includegraphics[width=\imgw]{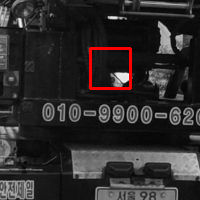} &
\includegraphics[width=\imgw]{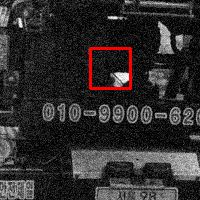} &
\includegraphics[width=\imgw]{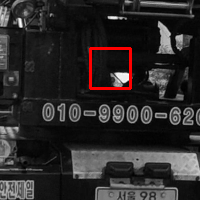} &
\includegraphics[width=\imgw]{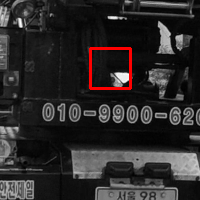} &
\includegraphics[width=\imgw]{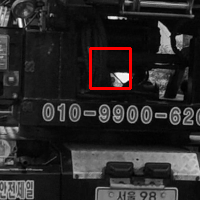} &
\includegraphics[width=\imgw]{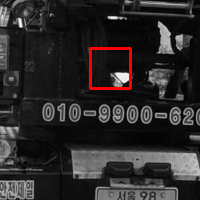} &
\includegraphics[width=\imgw]{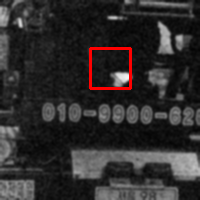} &
\includegraphics[width=\imgw]{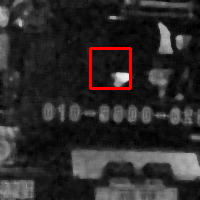} &
\includegraphics[width=\imgw]{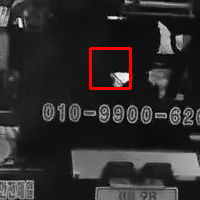} \\[0.6ex]

\includegraphics[width=\imgw]{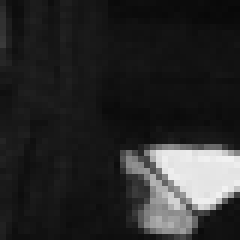} &
\includegraphics[width=\imgw]{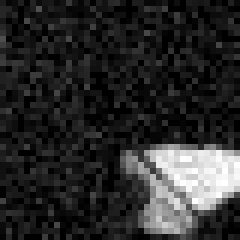} &
\includegraphics[width=\imgw]{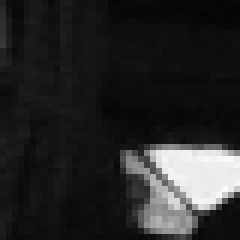} &
\includegraphics[width=\imgw]{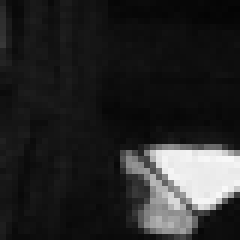} &
\includegraphics[width=\imgw]{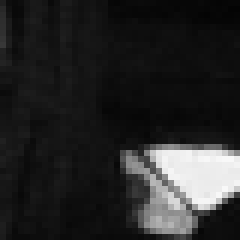} &
\includegraphics[width=\imgw]{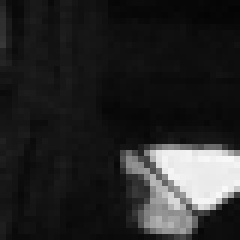} &
\includegraphics[width=\imgw]{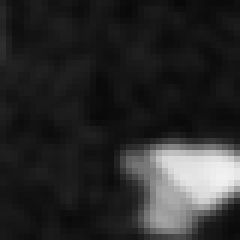} &
\includegraphics[width=\imgw]{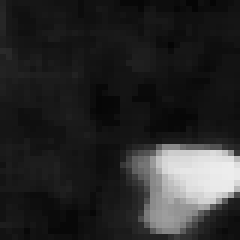} &
\includegraphics[width=\imgw]{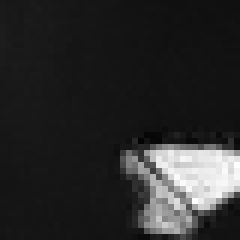} \\[1.2ex]

\includegraphics[width=\imgw]{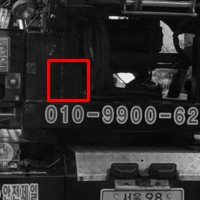} &
\includegraphics[width=\imgw]{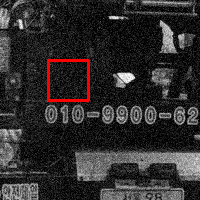} &
\includegraphics[width=\imgw]{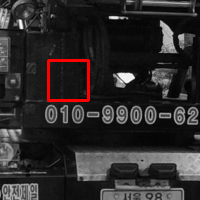} &
\includegraphics[width=\imgw]{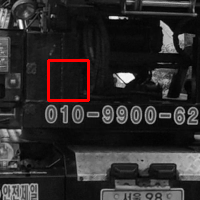} &
\includegraphics[width=\imgw]{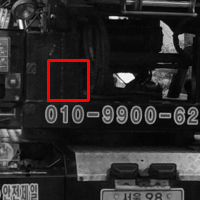} &
\includegraphics[width=\imgw]{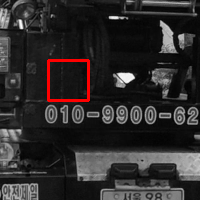} &
\includegraphics[width=\imgw]{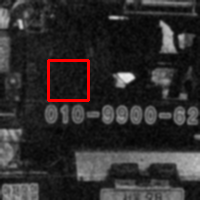} &
\includegraphics[width=\imgw]{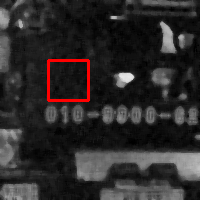} &
\includegraphics[width=\imgw]{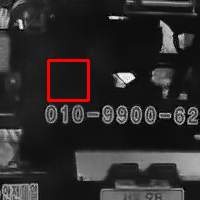} \\[0.6ex]

\includegraphics[width=\imgw]{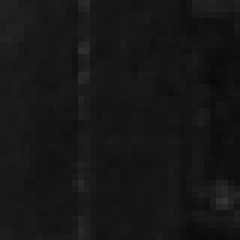} &
\includegraphics[width=\imgw]{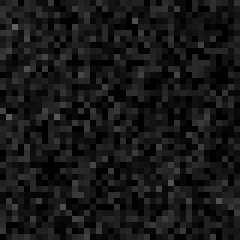} &
\includegraphics[width=\imgw]{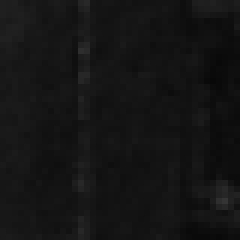} &
\includegraphics[width=\imgw]{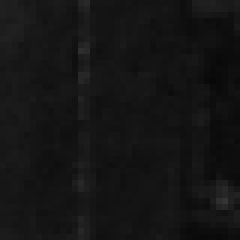} &
\includegraphics[width=\imgw]{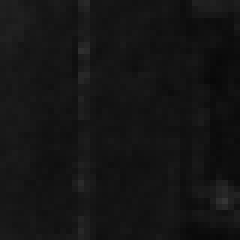} &
\includegraphics[width=\imgw]{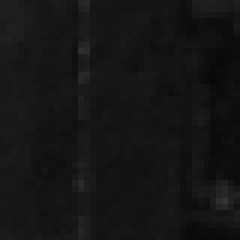} &
\includegraphics[width=\imgw]{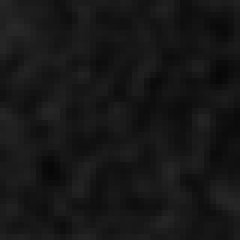} &
\includegraphics[width=\imgw]{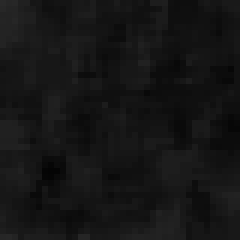} &
\includegraphics[width=\imgw]{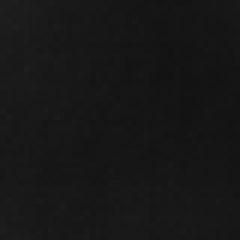} \\[1.2ex]

\includegraphics[width=\imgw]{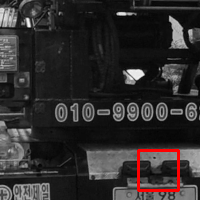} &
\includegraphics[width=\imgw]{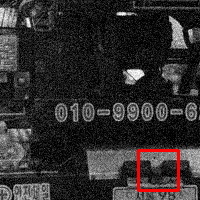} &
\includegraphics[width=\imgw]{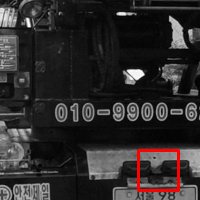} &
\includegraphics[width=\imgw]{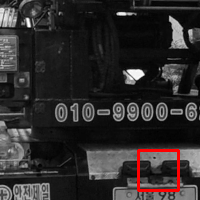} &
\includegraphics[width=\imgw]{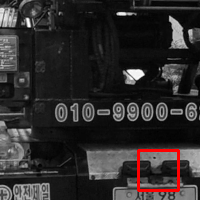} &
\includegraphics[width=\imgw]{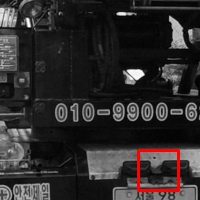} &
\includegraphics[width=\imgw]{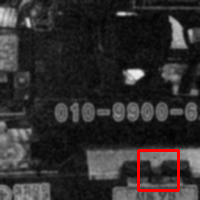} &
\includegraphics[width=\imgw]{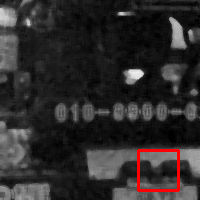} &
\includegraphics[width=\imgw]{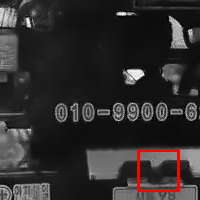} \\[0.6ex]

\includegraphics[width=\imgw]{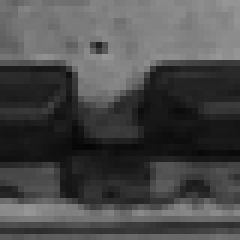} &
\includegraphics[width=\imgw]{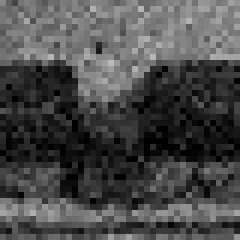} &
\includegraphics[width=\imgw]{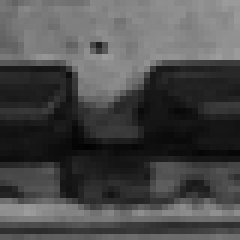} &
\includegraphics[width=\imgw]{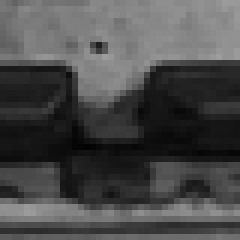} &
\includegraphics[width=\imgw]{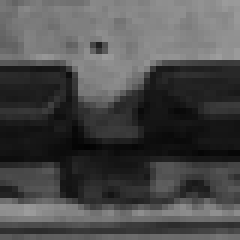} &
\includegraphics[width=\imgw]{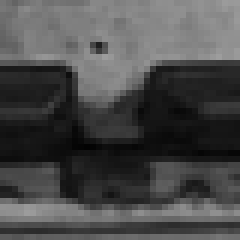} &
\includegraphics[width=\imgw]{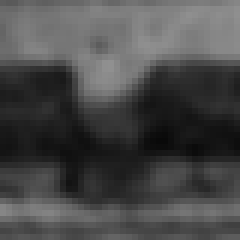} &
\includegraphics[width=\imgw]{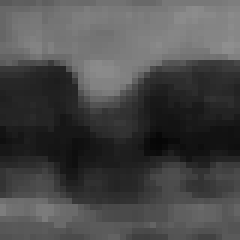} &
\includegraphics[width=\imgw]{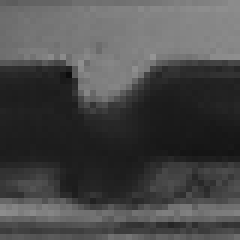} \\[1.2ex]

\scriptsize Original & \scriptsize Noisy & \scriptsize 2D-GFRFT & \scriptsize 2D-GBFRFT & \scriptsize JFRFT & \scriptsize GC-GFRFT & \scriptsize Gaussian & \scriptsize Median & \scriptsize NLM
\end{tabular}

\caption{Visual comparison on the REDS sequence for dynamic image denoising. For each frame, the first row shows the selected region on the full image, and the second row shows the corresponding magnified patch.}
\label{fig:reds_zoom_compare}
\end{figure*}

Fig.~\ref{fig:reds_zoom_compare} provides qualitative comparisons for dynamic image denoising. For each frame, a representative region is highlighted on the full image, and the corresponding magnified patch is shown beneath it. Compared with classical spatial filters, the fractional transform-based methods preserve more meaningful local structures while suppressing noise more effectively. Among them, GC-GFRFT produces cleaner textures and fewer residual noise artifacts, while avoiding the oversmoothing effects observed in Gaussian filtering, median filtering, and NLM. These observations are consistent with the quantitative results in Tables~\ref{tab:denoising_results_single} and~\ref{tab:denoising_results_avg}, further supporting the effectiveness of GC-GFRFT for dynamic image denoising.

\begin{table}[htbp]
	\centering
	\caption{Comparison of Denoising Performance (MSE/PSNR/SSIM) on the REDS Dataset}
	\label{tab:denoising_results_single}
	\resizebox{\columnwidth}{!}{
		\begin{tabular}{cccccc}
			\toprule
			Dataset & Method & Frame & MSE & PSNR & SSIM \\
			\midrule
		
			\multirow{21}{*}[-4.2ex]{REDS}
			& \multirow{3}{*}{2D-GFRFT}
			& Frame 1 & 10.9443 & 37.74 & 0.9791 \\
			& & Frame 2 & 4.2729 & 41.82 & 0.9947 \\
			& & Frame 3 & 8.3665 & 38.91 & 0.9889 \\
			\cmidrule(lr){2-6}
			
			& \multirow{3}{*}{2D-GBFRFT}
			& Frame 1 & 10.7856 & 37.80 & 0.9793 \\
			& & Frame 2 & 4.2707 & 41.83 & 0.9947 \\
			& & Frame 3 & 8.3324 & 38.92 & 0.9890 \\
			\cmidrule(lr){2-6}
			
			& \multirow{3}{*}{JFRFT}
			& Frame 1 & 10.6420 & 37.86 & 0.9795 \\
			& & Frame 2 & 4.2716 & 41.82 & 0.9947 \\
			& & Frame 3 & \textbf{8.2394} & \textbf{38.97} & \textbf{0.9891} \\
			\cmidrule(lr){2-6}
			
			& \multirow{3}{*}{GC-GFRFT}
			& Frame 1 & \textbf{10.4633} & \textbf{37.93} & \textbf{0.9796} \\
			& & Frame 2 & \textbf{4.1533} & \textbf{41.95} & \textbf{0.9948} \\
			& & Frame 3 & 8.3356 & 38.92 & 0.9890 \\
			\cmidrule(lr){2-6}
			
			& \multirow{3}{*}{Gaussian}
			& Frame 1 & 162.5462 & 26.02 & 0.7850 \\
			& & Frame 2 & 158.1820 & 26.14 & 0.7877 \\
			& & Frame 3 & 161.3213 & 26.05 & 0.7829 \\
			\cmidrule(lr){2-6}
			
			& \multirow{3}{*}{Median}
			& Frame 1 & 272.5740 & 23.78 & 0.6850 \\
			& & Frame 2 & 268.2410 & 23.85 & 0.6810 \\
			& & Frame 3 & 279.8108 & 23.66 & 0.6703 \\
			\cmidrule(lr){2-6}
			
			& \multirow{3}{*}{NLM}
			& Frame 1 & 89.5665 & 28.61 & 0.8332 \\
			& & Frame 2 & 89.2881 & 28.62 & 0.8302 \\
			& & Frame 3 & 93.9571 & 28.40 & 0.8272 \\
			\bottomrule
	\end{tabular}}
\end{table}

\begin{table}[htbp]
\centering
\caption{Denoising Performance in Terms of $MSE_{\text{avg}}$, $PSNR_{\text{avg}}$, and $SSIM_{\text{avg}}$ on the REDS Dataset}
\label{tab:denoising_results_avg}
\resizebox{\columnwidth}{!}{
\begin{tabular}{ccccc}
\toprule
Dataset & Method & $MSE_{\text{avg}}$ & $PSNR_{\text{avg}}$ & $SSIM_{\text{avg}}$ \\
\midrule
\multirow{7}{*}{REDS}
& 2D-GFRFT   & 7.8612 & 39.49 & 0.9876 \\
& 2D-GBFRFT  & 7.7963 & 39.52 & 0.9877 \\
& JFRFT      & 7.7177 & 39.55 & 0.9877 \\
& GC-GFRFT   & \textbf{7.6507} & \textbf{39.60} & \textbf{0.9878} \\
& Gaussian   & 160.6831 & 26.07 & 0.7852 \\
& Median     & 273.5419 & 23.76 & 0.6787 \\
& NLM        & 90.9372 & 28.54 & 0.8302 \\
\bottomrule

\end{tabular}}
\end{table}

\section{Conclusion and Future Work}
\label{sec:conclusion}

In this paper, we developed a unified fractional spectral framework for graph signals on Cartesian product graphs, extending conventional integer-order and single-parameter fractional transforms. We first proposed the 2D-GBFRFT, which assigns independent fractional orders to the factor graphs and relaxes the shared-order constraint of the standard 2D-GFRFT, thereby improving modeling flexibility for spatiotemporal signals. To bridge the graph-induced temporal basis and the discrete temporal basis, we further introduced the GC-GFRFT by constructing a coupling path on the unitary manifold. This construction preserves unitarity and admits a closed-form inverse, providing a principled link between connectivity-induced graph spectra and classical temporal spectral representations.

Building upon these transforms, we formulated a differentiable Wiener filtering approach in which the fractional orders are optimized end-to-end via gradient-based learning, while the coupling parameter serves as a fixed structural regularizer. Experiments on real-world tasks, including time-varying graph-signal denoising and dynamic image restoration, demonstrate consistent gains over existing fractional transforms and remain competitive with representative learning-based baselines.

Despite these promising results, the current framework still has several limitations. In particular, the proposed GC-GFRFT is presently restricted to interpolation between two endpoint unitary bases and therefore does not yet support more general spectral extrapolation. In addition, the coupling parameter $\lambda$ is treated as a fixed structural hyperparameter rather than being jointly learned with the remaining model parameters. Future work will therefore focus on extending the geodesic coupling mechanism beyond the current interpolation range to support spectral extrapolation. Another important direction is to integrate differentiable GC-GFRFT modules into deep graph neural networks as lightweight spectral layers, thereby enhancing spectral expressiveness with modest parameter overhead.

\appendices

\section{Proof of Properties~5--7}
\label{app:proof_properties_5_7}

\begin{IEEEproof}[\hspace{-0.8em}Proof of Property 5]
	From \eqref{eq:gc_Ft}, $\mathbf{F}_{t,\mathrm{GC}}^{(0;\beta)}=\mathbf{F}_{G_2}^{\beta}$.
	At $\lambda=1$, Assumption~1 gives $\exp(\log(\mathbf{W}_t))=\mathbf{W}_t$, hence
	\begin{equation*}
		\mathbf{F}_{t,\mathrm{GC}}^{(1;\beta)}
		=\mathbf{F}_{G_2}^{\beta}\mathbf{W}_t
		=\mathbf{F}_{G_2}^{\beta}(\mathbf{F}_{G_2}^{\beta})^{\mathrm H}\mathbf{F}^{\beta}
		=\mathbf{F}^{\beta}.
	\end{equation*}
	Substituting into \eqref{eq:FGC_def} yields \eqref{eq:FGC_degeneracy}.
\end{IEEEproof}

\begin{IEEEproof}[\hspace{-0.8em}Proof of Property 6]
	Let $\mathbf{A}\triangleq \log(\mathbf{W}_t)$. Under Assumption~1, $\mathbf{A}^{\mathrm H}=-\mathbf{A}$, so $\exp(\lambda\mathbf{A})$ is unitary for any real $\lambda$, and in particular for $\lambda\in[0,1]$.
	Thus $\mathbf{F}_{t,\mathrm{GC}}^{(\lambda;\beta)}=\mathbf{F}_{G_2}^{\beta}\exp(\lambda\mathbf{A})$ is unitary.
	Finally, using $(\mathbf{P}\otimes\mathbf{Q})^{\mathrm H}(\mathbf{P}\otimes\mathbf{Q})=(\mathbf{P}^{\mathrm H}\mathbf{P})\otimes(\mathbf{Q}^{\mathrm H}\mathbf{Q})$ with \eqref{eq:FGC_def} gives \eqref{eq:FGC_unitary}.
\end{IEEEproof}

\begin{IEEEproof}[\hspace{-0.8em}Proof of Property 7]
	Since $(\mathbf{F}^{\beta})^{\mathrm H}\mathbf{F}_{G_2}^{\beta}=\mathbf{W}_t^{-1}$ and $\log(\mathbf{W}_t^{-1})=-\log(\mathbf{W}_t)$ under Assumption~1, we have
	\begin{equation*}
		\widetilde{\mathbf{F}}_{t,\mathrm{GC}}^{(\lambda;\beta)}
		=\mathbf{F}^{\beta}\exp\!\big(-\lambda\log(\mathbf{W}_t)\big).
	\end{equation*}
	Using $\mathbf{F}^{\beta}=\mathbf{F}_{G_2}^{\beta}\exp(\log(\mathbf{W}_t))$, it follows that
	\begin{equation*}
		\widetilde{\mathbf{F}}_{t,\mathrm{GC}}^{(\lambda;\beta)}
		=\mathbf{F}_{G_2}^{\beta}\exp\!\big((1-\lambda)\log(\mathbf{W}_t)\big)
		=\mathbf{F}_{t,\mathrm{GC}}^{(1-\lambda;\beta)}.
	\end{equation*}
	Tensoring with $\mathbf{F}_{G_1}^{\alpha}$ yields \eqref{eq:FGC_symmetry}.
\end{IEEEproof}

\section{Proof of Theorem~1}
\label{app:proof_thm2}

For clarity, we reuse the notation in Section~\ref{sec:gcgfrft} and write the learnable spatial order as
$\alpha$. Throughout this proof, the coupling parameter $\lambda$ is regarded as a fixed
hyperparameter and is not updated by GD.

Recall from \eqref{eq:gc_vec} that the global GC operator admits the Kronecker form
\begin{equation}
	\mathbf{F}_{\mathrm{GC}}^{(\lambda;\alpha,\beta)}
	=
	\mathbf{F}_{t,\mathrm{GC}}^{(\lambda;\beta)}\otimes \mathbf{F}_{G_1}^{\alpha},
	\label{eq:app_FGC_kron}
\end{equation}
where the temporal GC basis is given in \eqref{eq:gc_Ft} by
\begin{equation}
	\mathbf{F}_{t,\mathrm{GC}}^{(\lambda;\beta)}
	=
	\mathbf{F}_{G_2}^{\beta}\,
	\exp\!\big(\lambda\,\log(\mathbf{W}_t(\beta))\big),
	\quad
	\mathbf{W}_t(\beta)=\big(\mathbf{F}_{G_2}^{\beta}\big)^{\mathrm H}\mathbf{F}^{\beta}.
	\label{eq:app_FtGC}
\end{equation}
By Assumption~1, $-1\notin\sigma(\mathbf{W}_t(\beta))$, hence the principal matrix logarithm
$\log(\mathbf{W}_t(\beta))$ is well-defined and Fr\'echet differentiable, and so is
$\exp(\lambda\log(\mathbf{W}_t(\beta)))$.

\textit{Differentiability w.r.t.\ $\alpha$.}
Since $\mathbf{F}_{G_1}^{\alpha}$ is a differentiable fractional operator with respect to its order,
$\partial \mathbf{F}_{G_1}^{\alpha}/\partial \alpha$ exists.
Using \eqref{eq:app_FGC_kron} and the Kronecker-product rule (with $\mathbf{F}_{t,\mathrm{GC}}^{(\lambda;\beta)}$
independent of $\alpha$), we obtain
\begin{equation}
	\frac{\partial \mathbf{F}_{\mathrm{GC}}^{(\lambda;\alpha,\beta)}}{\partial \alpha}
	=
	\mathbf{F}_{t,\mathrm{GC}}^{(\lambda;\beta)}\otimes
	\frac{\partial \mathbf{F}_{G_1}^{\alpha}}{\partial \alpha}.
	\label{eq:app_dFGC_alpha}
\end{equation}
Therefore, $\mathbf{F}_{\mathrm{GC}}^{(\lambda;\alpha,\beta)}$ is differentiable with respect to $\alpha$.

\textit{Differentiability w.r.t.\ $\beta$.}
From \eqref{eq:app_FtGC}, $\mathbf{F}_{t,\mathrm{GC}}^{(\lambda;\beta)}$ is a product of
$\mathbf{F}_{G_2}^{\beta}$ and $\exp(\lambda\log(\mathbf{W}_t(\beta)))$.
Applying the product rule and the chain rule for Fr\'echet derivatives yields
\begin{equation}
	\begin{aligned}
		\frac{\partial \mathbf{F}_{t,\mathrm{GC}}^{(\lambda;\beta)}}{\partial \beta}
		&=
		\frac{\partial \mathbf{F}_{G_2}^{\beta}}{\partial \beta}\,
		\exp\!\big(\lambda\log(\mathbf{W}_t(\beta))\big) \\
		&\quad
		+ \mathbf{F}_{G_2}^{\beta}\,
		L_{\exp}\!\big(\lambda\log(\mathbf{W}_t(\beta));\,\mathbf{E}_t(\beta)\big),
	\end{aligned}
	\label{eq:app_dFt_beta}
\end{equation}

\begin{equation}
	\mathbf{E}_t(\beta)
	\triangleq
	\lambda\,L_{\log}\!\big(\mathbf{W}_t(\beta);\,\mathbf{W}_t'(\beta)\big).
	\label{eq:app_Et_def}
\end{equation}
where $L_{\exp}(\cdot;\cdot)$ and $L_{\log}(\cdot;\cdot)$ denote the Fr\'echet derivatives of the matrix exponential
and principal logarithm, respectively. Moreover, differentiating $\mathbf{W}_t(\beta)=\big(\mathbf{F}_{G_2}^{\beta}\big)^{\mathrm H}\mathbf{F}^{\beta}$ in \eqref{eq:app_FtGC} gives
\begin{equation}
	\mathbf{W}_t'(\beta)
	=
	\left(\frac{\partial \mathbf{F}_{G_2}^{\beta}}{\partial \beta}\right)^{\!\mathrm H}\mathbf{F}^{\beta}
	+
	\big(\mathbf{F}_{G_2}^{\beta}\big)^{\mathrm H}\frac{\partial \mathbf{F}^{\beta}}{\partial \beta},
	\label{eq:app_dWt}
\end{equation}
and $\partial \mathbf{F}^{\beta}/\partial \beta$ exists since $\mathbf{F}^{\beta}$ is also a differentiable
fractional transform with respect to $\beta$. Hence
$\partial \mathbf{F}_{t,\mathrm{GC}}^{(\lambda;\beta)}/\partial \beta$ exists.

Finally, by \eqref{eq:app_FGC_kron},
\begin{equation}
	\frac{\partial \mathbf{F}_{\mathrm{GC}}^{(\lambda;\alpha,\beta)}}{\partial \beta}
	=
	\frac{\partial \mathbf{F}_{t,\mathrm{GC}}^{(\lambda;\beta)}}{\partial \beta}
	\otimes \mathbf{F}_{G_1}^{\alpha},
	\label{eq:app_dFGC_beta}
\end{equation}
which establishes differentiability with respect to $\beta$.

\textit{Differentiability of the objective.}
By Property~6 in Section~\ref{sec:gcgfrft}, $\mathbf{F}_{\mathrm{GC}}^{(\lambda;\alpha,\beta)}$ is unitary for $\lambda\in[0,1]$,
so its inverse exists and satisfies
$(\mathbf{F}_{\mathrm{GC}}^{(\lambda;\alpha,\beta)})^{-1}=(\mathbf{F}_{\mathrm{GC}}^{(\lambda;\alpha,\beta)})^{\mathrm H}$.
Therefore, the mapping
\[
\mathbf{y}\mapsto
(\mathbf{F}_{\mathrm{GC}}^{(\lambda;\alpha,\beta)})^{-1}\mathbf{H}_{\mathrm{GC}}
\mathbf{F}_{\mathrm{GC}}^{(\lambda;\alpha,\beta)}\mathbf{y}
\]
is a composition of differentiable matrix functions and linear operators in $(\alpha,\beta,\mathbf{h})$.
Consequently, the loss in \eqref{eq:obj-gc} is differentiable with respect to $(\alpha,\beta,\mathbf{h})$ for fixed admissible $\lambda$.
\hfill$\blacksquare$


\bibliographystyle{IEEEtran}
\bibliography{reference}

\newpage

\vfill

\end{document}